\documentclass[sigconf,pdfusetitle,pdfdisplaydoctitle]{acmart}
\AtBeginDocument{%
  \providecommand\BibTeX{{%
    \normalfont B\kern-0.5em{\scshape i\kern-0.25em b}\kern-0.8em\TeX}}}

\usepackage{enumerate}
\usepackage{graphicx}
\usepackage{subcaption}
\usepackage{float}
\usepackage{multirow}
\usepackage{array}
\usepackage{makecell}
\usepackage{gensymb}
\usepackage{enumitem}
\usepackage[skip=3pt]{caption}  
\definecolor{lightgray}{gray}{0.95}  

\usepackage{tikz}
\usetikzlibrary{arrows.meta, positioning}
\definecolor{ovgray}{HTML}{5F5E5A}
\definecolor{ovteal}{HTML}{0F6E56}
\definecolor{ovpurple}{HTML}{534AB7}

\copyrightyear{2026}
\acmYear{2026}
\setcopyright{cc}
\setcctype{by}
\acmConference[ASSETS '26]{The 28th International ACM SIGACCESS Conference on Computers and Accessibility}{October 25--28, 2026}{Vila Nova de Gaia, Portugal}
\acmBooktitle{The 28th International ACM SIGACCESS Conference on Computers and Accessibility (ASSETS '26), October 25--28, 2026, Vila Nova de Gaia, Portugal}
\acmDOI{10.1145/3797867.3828974}
\acmISBN{979-8-4007-2521-0/2026/10}

\newcommand{\system}{SceneGlance}

\definecolor{brownishred}{RGB}{178, 34, 34}

\DeclareRobustCommand{\colorchange}[1]{%
  #1%
}

\begin{document}

\title[SceneGlance]{What to Distinguish and How? Opportunities and Challenges of Augmenting Multiple, Cluttered Objects in Complex Scenes for People with Low Vision}

\settopmatter{authorsperrow=3}

\author{Yuheng Wu}
\orcid{0009-0005-1828-400X}
\affiliation{%
  \department{Department of Computer Sciences}
  \institution{University of Wisconsin-Madison}
  \city{Madison}
  \state{WI}
  \country{USA}
}
\email{yuheng.wu@wisc.edu}

\author{Ruijia Chen}
\orcid{0000-0002-1655-6228}
\affiliation{%
  \department{Department of Computer Sciences}
  \institution{University of Wisconsin-Madison}
  \city{Madison}
  \state{WI}
  \country{USA}
}
\email{ruijia.chen@wisc.edu}

\author{Jaewook Lee}
\orcid{0000-0002-1481-9290}
\affiliation{%
  \department{Paul G. Allen School of Computer Science \& Engineering}
  \institution{University of Washington}
  \city{Seattle}
  \state{WA}
  \country{USA}
}
\email{jaewook4@cs.washington.edu}

\author{Jia Li}
\orcid{0009-0008-0198-6699}
\affiliation{%
  \department{Department of Computer Science}
  \institution{The University of Texas at Dallas}
  \city{Richardson}
  \state{TX}
  \country{USA}
}
\email{jia.li@utdallas.edu}

\author{Kexin Zhang}
\orcid{0009-0009-4078-8780}
\affiliation{%
  \department{Department of Computer Sciences}
  \institution{University of Wisconsin-Madison}
  \city{Madison}
  \state{WI}
  \country{USA}
}
\email{kzhang284@wisc.edu}

\author{Meng Fong Lio}
\orcid{0009-0007-8692-6670}
\affiliation{%
  \department{Department of Computer Sciences}
  \institution{University of Wisconsin-Madison}
  \city{Madison}
  \state{WI}
  \country{USA}
}
\email{mengfong@wisc.edu}

\author{Weibing Wang}
\orcid{0009-0005-4714-3142}
\affiliation{%
  \department{Department of Computer Sciences}
  \institution{University of Wisconsin-Madison}
  \city{Madison}
  \state{WI}
  \country{USA}
}
\email{wwang652@wisc.edu}

\author{Sanbrita Mondal}
\orcid{0000-0003-4454-8978}
\affiliation{%
  \department{School of Medicine and Public Health}
  \institution{University of Wisconsin-Madison}
  \city{Madison}
  \state{WI}
  \country{USA}
}
\email{smondal4@wisc.edu}

\author{Jon E. Froehlich}
\orcid{0000-0001-8291-3353}
\affiliation{%
  \department{Paul G. Allen School of Computer Science \& Engineering}
  \institution{University of Washington}
  \city{Seattle}
  \state{WA}
  \country{USA}
}
\email{jonf@cs.uw.edu}

\author{Yapeng Tian}
\orcid{0000-0003-1423-4513}
\affiliation{%
  \department{Department of Computer Science}
  \institution{The University of Texas at Dallas}
  \city{Richardson}
  \state{TX}
  \country{USA}
}
\email{yapeng.tian@utdallas.edu}

\author{Yuhang Zhao}
\orcid{0000-0003-3686-695X}
\affiliation{%
  \department{Department of Computer Sciences}
  \institution{University of Wisconsin-Madison}
  \city{Madison}
  \state{WI}
  \country{USA}
}
\email{yuhang.zhao@cs.wisc.edu}

\renewcommand{\shortauthors}{Wu et al.}

\begin{abstract}
People with low vision (PLV) struggle to perceive complex scenes like busy kitchens and crowded streets, which contain many objects, visual clutter, and dynamic elements. Prior AR systems for low vision either enhance low-level visual features or augment task-relevant objects for single tasks in simple settings, leaving multi-object augmentation in complex scenes underexplored. Informed by a formative study characterizing important objects and their perceived importance for PLV, we built \system{}, a wearable AR system that recognizes important objects and visually distinguishes them by importance level. Through a controlled lab study with 12 PLV in a mock-up kitchen scene and a free-form think-aloud study with 13 PLV navigating an outdoor route, we found that AR distinction on object importance shifted PLV's attention toward objects of higher importance, and supported perception strategies such as building mental snapshots from the augmentation distribution and hierarchical scanning by importance. \colorchange{However, this attention shift came with a tradeoff of reduced overall scene recall.} The studies also surfaced challenges posed by AR augmentations in complex scenes, such as adjacent augmentations blending or interfering with each other, yielding design implications for more practical AR vision enhancement systems in the complex real world.

\end{abstract}

\ccsdesc[500]{Human-centered computing~Accessibility systems and tools}
\ccsdesc[500]{Human-centered computing~Accessibility technologies}
\ccsdesc[500]{Human-centered computing~Mixed / augmented reality}

\keywords{Accessibility, Augmented Reality, Low Vision, Vision Enhancement}

\maketitle

\begin{figure*}[t]
    \centering
    \includegraphics[width=\textwidth]{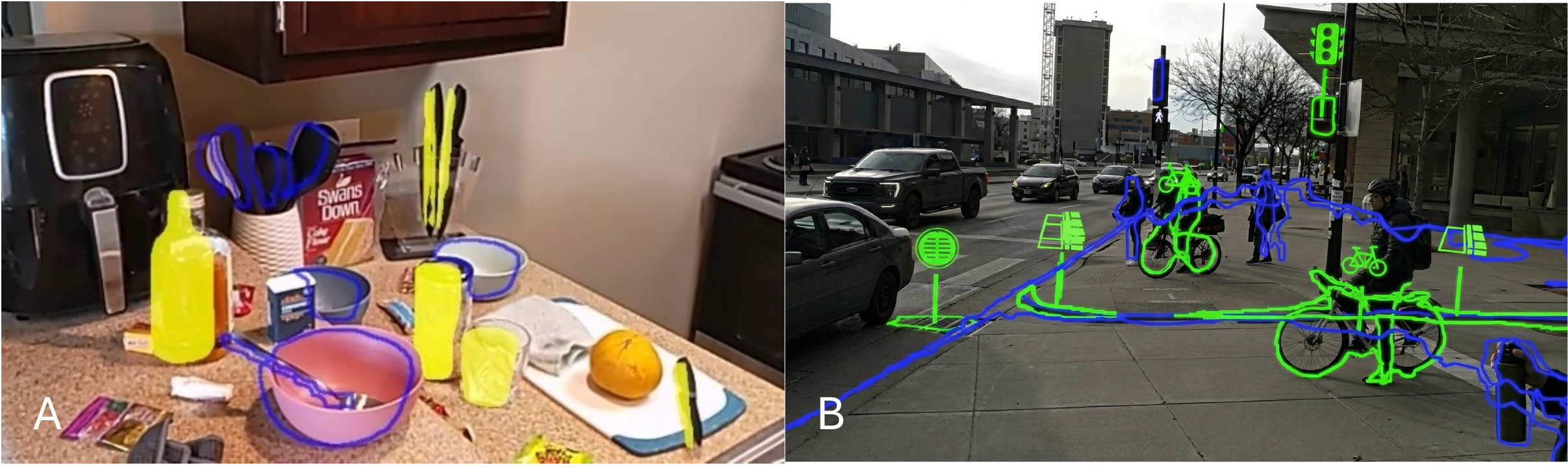}
  \caption{We explore the opportunities and design challenges of augmenting and distinguishing multiple objects in complex scenes to support scene perception for people with low vision, using SceneGlance as a technology probe. (A) SceneGlance distinguishes multiple objects in a complex kitchen, with primary-important objects (glasses, knives) overlaid in solid yellow, and secondary ones (bowls, spoons, ladles) in a blue outline. (B) SceneGlance distinguishes multiple objects on a crowded street, with primary-important objects (curb, sewer drain, pedestrian signal, bicycles) in a green outline with floating icon labels, and secondary ones (sidewalk, pedestrian) in a blue outline only.}
  \Description{This figure shows that \system{} augments and distinguishes multiple important objects in complex environments to support low vision people with scene perception. This figure has two subfigures. Subfigure A is \system{} augmenting multiple objects in a complex kitchen, with primary-important objects (glasses, knives) in yellow solid overlays and secondary-important objects (bowls, spoons, ladles) in blue outline. Subfigure B is \system{} augmenting multiple objects on a crowded street, with primary-important objects (curb, sewer drain, pedestrian signal, bicycles) in green outline with floating green icon labels, and secondary-important objects (sidewalk, pedestrian) in blue outline only.}
  \label{fig:teaser}
\end{figure*}

\section{Introduction}
Complex scenes---characterized by a high quantity and variety of objects, high visual clutter, complex spatial layout, and the presence of dynamic, unpredictable objects \cite{oliva2004identifying, chai2010scene, rosenholtz2007measuring, miniukovich2014quantification, kyle2023characterising, lee2021understanding}---are ubiquitous in everyday settings like a busy kitchen or crowded street. 
For people with low vision (PLV) who experience uncorrectable vision loss \cite{nihVisionNational}, these scenes pose significant challenges in daily tasks, such as finding products on crowded store shelves \cite{szpiro2016shopping, khattab2015understanding} or avoiding hazards on busy streets \cite{williams2013pray, muller2022traveling}.


Prior research has explored using augmented reality (AR) systems to enhance PLV's visual abilities \cite{zhao2017understanding}. 
Conventional AR systems enhance scene visibility by augmenting low-level visual features such as edges and contrast \cite{hwang2014augmented, zhao2015foresee}.
However, they also amplify details unnecessary for scene interpretation, which may create visual clutter and cognitive overload in complex environments \cite{everingham1999head, al2010designing}.
To address this, recent research has developed AR systems that assist PLV with specific tasks by augmenting task-relevant objects, such as a target product for grocery shopping \cite{zhao2016cuesee}, or graspable and hazardous parts of kitchen tools for cooking \cite{lee2024cookar}.
However, these systems were only evaluated in relatively simple settings with few augmented objects and low visual clutter.

In contrast, perceiving complex scenes requires quickly scanning and identifying multiple important objects simultaneously while filtering out distractors \cite{wolfe2021guided, henderson2019meaning}.
To cope with this challenge, the human visual system selectively allocates attention to a subset of important objects in the scene for more efficient perception \cite{wiesmann2023disentangling, macevoy2011constructing, henderson2019meaning, vo2021meaning}. However, this process can be challenging for PLV as their vision loss hinders fast identification of important objects. We seek to support this selective attention by automatically detecting and augmenting important objects. However, in complex scenes with many important objects, simply augmenting all of them equally can cause visual clutter and fail to provide guidance on where to look. Therefore, we propose \textbf{\textit{AR distinction}}, a method that visually distinguishes objects of different importance through different AR augmentations, thus enhancing important objects while reducing visual overwhelm in complex scenes. 

To understand what objects are important for PLV in complex scenes and how to augment and distinguish them, we conducted a formative study with six PLV in two complex scenarios: \textit{cooking in a busy kitchen} and \textit{navigating a crowded street} \cite{bilyk2009food, li2021cooking, szpiro2016shopping, muller2022traveling}. The study characterized three categories of important objects (safety-related, visually challenging, frequently used) and two factors determining their importance levels, and derived guidelines for AR augmentation and distinction. Informed by these findings, we developed \textbf{\textit{\system{}}}, a head-mounted AR system that combines RGB and depth information to recognize, locate, and augment multiple important objects in 3D space with AR distinction. To understand how \system{} supports PLV in complex scene perception and what challenges arise, we investigate:



\begin{itemize}
    \item[RQ1:] How do AR augmentation and distinction support and change PLV's perception of complex scenes?
    \item[RQ2:] What are the design challenges and implications when augmenting visually complex scenes with AR for PLV?
\end{itemize}

\begin{figure*}[t]
  \centering
  \includegraphics[width=\textwidth]{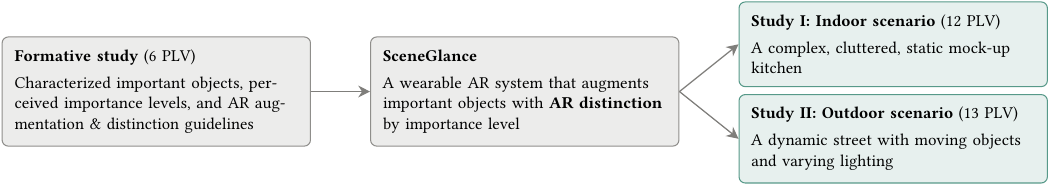}
  \caption{\colorchange{Research method overview. A formative study with six PLV characterized important objects, their perceived importance levels, and AR augmentation and distinction guidelines. The results informed the design of \system{}, a wearable AR system that augments important objects with AR distinction by importance level. We evaluated \system{} in two complementary scenarios: Study~I, a complex, cluttered, static indoor kitchen, and Study~II, a dynamic outdoor street, each examining how AR distinction supports complex scene perception and the design challenges arising in this context.}}
  \Description{This image showed an overview of the research method: A formative study with six PLV informs the design of SceneGlance, a wearable AR system that augments important objects with AR distinction by importance level. SceneGlance is evaluated in two scenarios: Study I, an indoor mock-up kitchen (12 participants), and Study II, an outdoor street (13 participants).}
  \label{fig:overview}
\end{figure*}

\colorchange{To answer these questions, we evaluated \system{} in two scenarios: an indoor mock-up kitchen representing a complex, cluttered, static scene, and an outdoor scenario that supplements it with dynamic objects, reflected in two studies (Figure~\ref{fig:overview}).}
We first conducted a well-controlled lab study with 12 PLV perceiving a mock-up kitchen countertop to understand the impact of AR distinction on PLV's viewing strategy and mental map development.
The study revealed that, compared to perceiving without augmentation, \system{} effectively shifted participants' attention toward objects of higher importance. Beyond attention shift, AR distinction helped PLV perceive complex scenes by enabling them to build a scene snapshot from augmentation distribution and providing visual guidance for scanning. 
\colorchange{However, augmenting many objects reduced overall object recall compared to perceiving without augmentation, revealing an attention-recall tradeoff from multi-object augmentation in dense, cluttered scenes}. Our study also revealed new perception challenges of applying AR augmentations to complex, cluttered scenes, such as partially occluded and visually similar objects, adjacent augmentations blending in, and spatial misalignment between objects and augmentations. 
We further conducted a free-form think-aloud study with 13 PLV navigating outdoor streets using \system{}, revealing unique challenges and needs, such as augmenting continuous ground surfaces (e.g., sidewalks) and dynamic objects (e.g., pedestrians, cyclists). Our studies also uncovered participants' changing preferences on AR distinction designs between indoor and outdoor environments due to lighting conditions. We finally derive design implications for future AR vision enhancement systems in the complex real world.


\section{Related Work}
Our work builds on prior research on the challenges of perceiving complex scenes for people with blindness or low vision, as well as AR-based visual augmentation systems for low vision.

\subsection{Challenges of Perceiving Complex Scenes for People with Blindness or Low Vision}
Many key daily activities require the ability to perceive complex scenes, such as locating ingredients in a crowded kitchen or walking down a bustling street.
A complex scene refers to an environment containing numerous objects with diverse visual features that increase the difficulty of perception and interpretation \cite{guan2022modelling, oliva2004identifying}. Prior work has characterized scene complexity as a multidimensional concept, influenced by the number and variety of objects \cite{oliva2004identifying, chai2010scene}, the degree of visual clutter \cite{rosenholtz2007measuring, miniukovich2014quantification, kyle2023characterising}, the presence and temporal unpredictability of dynamic objects \cite{lee2021understanding}, and the global scene layout, including symmetry, openness, and the spatial distribution of scene elements \cite{oliva2004identifying, kyle2023characterising}. 
These dimensions pose challenges for everyday visual activities even for sighted individuals, such as slowing visual search \cite{rosenholtz2007measuring, henderson2009influence, neider2011cutting, tadin2012peripheral}, impeding object recognition \cite{whitney2011visual, pelli2008uncrowded, wallace2017object}, and reducing scene recall \cite{alvarez2004capacity}.

Perceiving complex scenes requires recognizing multiple key objects and understanding their spatial relationships \cite{henderson2019meaning, vo2021meaning, wiesmann2023disentangling, macevoy2011constructing}. However, this is particularly challenging for PLV as scene complexity compounds their visual challenges. For example, visual clutter disproportionately impairs object recognition for people with central vision loss, as it is difficult for the peripheral visual system to clearly separate the target object's visual features from the surrounding distractors \cite{wallace2017object, thibaut2020object}. Moreover, for people with peripheral vision loss, navigating complex environments with many objects requires high mental effort, as they must intentionally decide where to direct their gaze without the visual preview afforded by peripheral vision \cite{turano1998mental}.
These challenges obstruct everyday activities, such as
navigating crowded store aisles or locating products among cluttered shelves during shopping \cite{szpiro2016shopping, khattab2015understanding, jones2019analysis, tullio2021you}, locating and distinguishing visually similar kitchen items in a complex kitchen \cite{bilyk2009food, wang2023characterizing, li2021cooking}, and noticing 
moving hazards (e.g., pedestrians, trolleys, cars) \cite{jeamwatthanachai2019indoor, williams2013pray, szpiro2016shopping} or identifying wayfinding cues (e.g., doors, intersections) \cite{jeamwatthanachai2019indoor, zeng2015survey, muller2022traveling} in busy navigation spaces.
These challenges significantly impact the independence, quality of life, and well-being of PLV \cite{jones2019analysis, teitelman2005psychosocial, desrosiers2009participation, haymes2002relationship, montero2005nutrition}.

To address these challenges, prior work has developed assistive technologies that support scene understanding for people with blindness or low vision through non-visual cues.
Some systems leverage object recognition models on users' camera feeds to recognize nearby objects and announce them via spatial audio \cite{dasila2017real, eckert2018object, kommey2019smart, killough_vrsight_2025}. However, in complex scenes with many objects, announcing each individual object can cause information overload.
Instead of simply enumerating objects, more recent work uses multi-modal large language models (MLLMs) to interpret users' real-time camera streams and generate high-level
scene descriptions \cite{magay2024light, hao2024chatmap, seeingaiSeeingTalking, gonzalez2024investigating, rao2021google, hao2024multi, kuribayashi2025wanderguide, chang2024worldscribe, mathis2025lifeinsight, chen2026navinote}.
To further reduce cognitive overload in complex environments, VIPTour \cite{lin2025ai} distills object information based on aesthetics, freshness, and basic needs into a simplified hierarchical representation that users can selectively explore on smartphones via gestures. NaviNote \cite{chen2026navinote} similarly lets users access location-based audio annotations of their surroundings on demand to avoid information overload.

Despite the potential of supporting complex scene perception, existing tools focus on non-visual feedback for blind users, overlooking the needs of PLV who still rely heavily on their residual vision for daily activities and prefer visual enhancement tools over audio cues \cite{szpiro2016shopping, szpiro2016people}. 
It remains unexplored how to enhance PLV's visual perception of complex environments with a large number of cluttered objects (sometimes visually similar or even moving) while minimizing visual and cognitive load. To fill this gap, we investigate the unique design opportunities and challenges of AR-based visual augmentations on complex scenes for people with low vision.




\subsection{AR-based Visual Augmentation Systems for Low Vision}
AR systems can support PLV by directly enhancing users' vision through real-time visual enhancements \cite{zhao2017understanding}. Conventional AR systems enhance scene visibility by augmenting the user's full field of view using image processing techniques, such as magnification \cite{zhao2019designing, stearns2018reading, zhao2015foresee}, edge enhancement \cite{hwang2014augmented, kwon2012contour, zhao2015foresee}, contrast enhancement \cite{zhao2015foresee}, scene recoloring based on depth information \cite{angelopoulos2019enhanced, hicks2013depth, van2015improving}, and pixel remapping for field of view loss \cite{sayed2020mobility, Sadeghzadeh2024ARVA, luo2006use, zhao2019computational}.
For example, Hwang et al. \cite{hwang2014augmented} designed an edge enhancement system on Google Glass that increases edge contrast to help people with age-related macular degeneration view their environments. Zhao et al. developed \textit{ForeSee} \cite{zhao2015foresee}, a head-mounted system rendering magnification, edge enhancement, and contrast enhancement to support PLV with diverse visual conditions.
While promising, these systems augment all low-level visual features equally, including details irrelevant to the user's task, potentially increasing visual clutter and distraction in complex scenes \cite{al2010designing, everingham1999head}. For example, in a complex outdoor scene, an edge enhancement system may improve the visibility of cars by enhancing their outer edges, but it also highlights irrelevant details like cloud texture and road surface, causing visual overload \cite{everingham1999head}.

More recent AR systems support PLV in specific daily activities by selectively augmenting task-relevant objects with visual cues \cite{zhao2016cuesee, lee2024cookar, dylan2023arnavigation, huang2019augmented, lang2021pressing, lee2024towards, chen2025visimark, zhao2019arstairs, gamage2025smart}.
For example, Zhao et al.'s \textit{CueSee} \cite{zhao2016cuesee} facilitates visual search by enhancing the target object with five types of visual augmentations (e.g., a flashing outline on the object, a guideline pointing at the object);
Lee et al.'s \textit{CookAR} \cite{lee2024cookar} supports kitchen tool interaction by rendering green overlays on graspable areas (e.g., knife handle) and red overlays on hazardous areas (e.g., blade) to indicate tool affordances;
Lee et al.'s \textit{ARSports} \cite{lee2024towards} assists PLV in playing tennis and basketball by rendering colored overlays on sport-related objects (e.g., basketball, tennis racket, player) to help detect and locate them faster; and
Chen et al.'s \textit{VisiMark} \cite{chen2025visimark} supports indoor navigation by rendering icon labels at landmarks along the route and signboard overviews at intersections to help PLV identify landmarks and preview hallway structure.
\colorchange{Beyond ocular low vision, Gamage et al. \cite{gamage2025smart} extended AR augmentation to cerebral visual impairment (CVI), co-designing augmentations with two CVI adults for daily difficulties such as locating a target in cluttered scenes and guiding attention toward salient objects.}

While these systems augment only task-relevant objects instead of the entire scene to reduce visual distraction, systems like \textit{CookAR} \cite{lee2024cookar} and \textit{ARSports} \cite{lee2024towards} were evaluated in relatively simple settings with a small number of augmented objects and low visual clutter. \textit{CueSee} \cite{zhao2016cuesee} was similarly evaluated on a mock grocery store shelf with a structured layout (e.g., products sequentially placed with no occlusion or clutter) and only one augmented target product at a time.
\textit{VisiMark} \cite{chen2025visimark} augmented more than one landmark along indoor hallways, but it was evaluated using simple routes with only 4--5 landmarks per route, and the landmarks were far from each other without introducing visual crowdedness.


In complex scenes, new challenges may arise for AR systems, such as visual clutter from augmentations \cite{tatzgern2016adaptive} and occlusion between augmentations and real-world objects \cite{park2025exploring}, which may pose additional challenges for PLV due to their visual impairments. However, to our knowledge, no research has thoroughly investigated how AR-based visual augmentations can support complex scene interpretation.
Our paper seeks to address this problem by exploring different AR object augmentation and distinction techniques in visually complex, cluttered scenes for PLV, revealing unique design challenges and implications for this context. 
\section{Formative Study}
To inform the design of \system{}, we first conducted a formative study with six PLV to understand: (1) objects that are important for PLV to perceive in complex scenes; (2) the importance level of these objects (i.e., whether all objects are equally important, and which objects are more important); and (3) PLV's design preferences for augmenting and distinguishing different important objects with AR in complex scenes.

\subsection{Method}

\subsubsection{Participants} We recruited six participants (P1--P6; age: $M=66, SD=13.83$; two male, four female) through contact lists from local low vision clinics and communities. Participants were eligible if they were at least 18 years old, had low vision, and used functional vision in daily tasks (e.g., using magnifying glasses to read). Table~\ref{tab:demographics_formative} in Appendix~\ref{app:demographic_formative} shows their demographics, visual conditions, and prior AR experience. Participants covered a broad range of low vision conditions, such as central vision loss (P1, P6), peripheral vision loss (P2-P5), and severe low visual acuity (P5). Three participants (P2, P3, P5) were legally blind (i.e., could not read the 20/100 line after best correction, or had field of view narrower than 20 degrees) \cite{aoaLegalBlindness}.
Two participants (P2, P5) had prior AR experience but did not use it regularly. Participants were compensated \$25 per hour. This study was approved by our university IRB.

\subsubsection{Procedure}
We conducted a single-session two-hour interview and observational study at each participant's residence to understand their daily experience. We focused on two representative scenarios, \textit{cooking in a busy kitchen} and \textit{navigating a crowded street}, as complex, challenging indoor and outdoor tasks, respectively \cite{remillard2024everyday}. 
We began the study by asking about participants' demographic information, visual conditions, and prior AR experience. Then, we demonstrated a HoloLens-based AR prototype that recognized a water bottle and augmented it with bright yellow outlines to help participants fully understand the concept of \textit{object augmentation}.

\textbf{Understanding Important Objects and Importance Level.} With a better understanding of AR augmentations, we discussed important objects and their importance levels in kitchen and street environments. In the kitchen scenario, participants first showed us their kitchens and described all objects they considered important to augment and explained their rationales.
We further probed them by showing an egocentric video\footnote{\url{https://drive.google.com/file/d/15MMxi4EwVKo-6bSzYB-ExHGP042ZR5T9/view?usp=sharing}} of various cooking tasks (e.g., cutting, transferring food) in complex kitchen environments  \cite{kashyap2020cooking, li2021cooking, bilyk2009food}, extracted from the Epic-Kitchens-100 dataset \cite{Damen2018EPICKITCHENS}. While watching the video, participants were asked to imagine themselves cooking and ``think aloud,'' describing what objects they considered important to augment and why. The video was played on a large display, and participants could pause and zoom in to examine certain frames as needed.
Finally, participants rated the importance level of each important object they mentioned on a five-point scale (1 means least important for augmentation; 5 means most important for augmentation) and explained their rationales.

For street navigation, we shadowed participants walking along two blocks in their neighborhoods. The routes were pre-selected 
to ensure they included sufficient potentially important or challenging objects (e.g., curb, pole, traffic sign) \cite{starke2020everyday, zeng2015survey}.
During the navigation, participants ``thought aloud'' to describe important objects they encountered and why. After the walk, we further probed them with an egocentric video of navigating on a busy street\footnote{\url{https://drive.google.com/file/d/1ASs8-ER3fscb\_8lhTyX9ejRliIxCwr9e/view?usp=sharing}} and asked them to point out objects they preferred to augment. Participants then reviewed their list of important objects for street navigation and rated importance levels for each object on the same five-point scale.

\textbf{AR Augmentation and Distinction Co-Design.} Finally, we conducted a probe-based co-design to understand participants' preferences for augmenting and distinguishing objects of different importance. We probed participants using slides by presenting four base augmentations for individual objects, adapted from prior research on low vision augmentation \cite{zhao2016cuesee, lee2024cookar, chen2025visimark}: (1) a static outline around each object, (2) a solid overlay, (3) a flashing outline, and (4) a text label floating above each object with its name (Figure \ref{fig:design_probe_1} in Appendix~\ref{sec:appendix_formative_probe}).
We illustrated each design on two kitchen example images and two street view example images to provide participants with sufficient context. Participants commented on the effectiveness, preferences, and suitable scenarios for each design and brainstormed other design ideas.

We then investigated how participants prefer to distinguish objects of different importance levels. We probed with four AR distinction methods (i.e., using different AR augmentation designs to distinguish different objects): (1) by form---solid overlay for high-importance objects vs. static outline for low-importance objects; (2) by color---bright yellow outline vs. dark blue outline; (3) by visual effect---flashing outline vs. static outline; and (4) by additional visual information---static outline and text label combined vs. static outline only  (Figure \ref{fig:design_probe_2} in Appendix~\ref{sec:appendix_formative_probe}). 
Participants viewed these AR distinction designs on our example images, critiqued each method, and brainstormed preferred design ideas.

Following thematic analysis \cite{braun2006using}, two researchers independently open-coded two transcripts (33\% of the data) to develop an initial codebook upon agreement. One researcher coded the remaining transcripts and derived themes on the characteristics of important objects, importance levels, and preferred AR augmentation and distinction designs.

\subsection{Findings}
\label{subsec:formative_findings}
Our formative study characterized important objects, factors influencing their importance levels, and design guidelines for augmenting and distinguishing multiple objects in complex scenes, \colorchange{extending prior work that identified important objects for PLV without distinguishing their perceived importance \cite{islam2024identifying}}.

\subsubsection{Characterizing Important Objects}
\label{subsubsec:salient_objects}
We identified three categories of important objects: \textit{safety-related}, \textit{visually challenging}, and \textit{frequently used} objects.

\textbf{Safety-related objects}.
All participants considered safety-related objects as important, including both hazards and risk indicators.
Hazards included \textit{tripping hazards} such as curbs and sewer drains (P1--P6), 
\textit{cutting hazards} such as knives and scissors (P1, P2, P5, P6), and \textit{burning hazards} such as stovetop surfaces and carafes (P1--P6). Beyond hazards, all participants found objects that indicated potential risks or safety boundaries to be important, such as the indicator lights on stoves (P1), crosswalks (P1, P2, P4–P6), pedestrian signals (P1--P6), 
and sidewalk boundaries (P1, P2, P4-P6). 

\textbf{Visually challenging objects}. All participants considered visually challenging objects as important to augment, including low-contrast objects (P1--P6; e.g., a black spatula on a dark counter, unclear crosswalk lines), transparent objects (P1--P6; e.g., glasses), cluttered objects (P1--P4, P6; e.g., a stack of dishes), objects of similar shape (P2, P3, P5, P6; e.g., spoons and forks), and small objects (P1, P3, P5; e.g., small clips, pedestrian signals).


\textbf{Frequently used objects}.
Five participants (P1--P3, P5, P6) considered frequently used objects to be important to augment, such as light switches (P3), utensils in the kitchen (P2), bus stops (P1, P3, P5), and traffic signs (P1--P3, P5).


\subsubsection{Factors that Impact Object Importance Levels}
\label{subsubsec:formative_priority}
We identified two factors influencing the perceived importance level of objects, including risk severity and visual difficulty.

\textbf{Risk Severity}.
All participants considered objects that posed higher safety risks to be more important to augment. Cutting and burning hazards were typically rated as high-importance due to the potential for severe injuries (P1--P6). For tripping hazards, lower objects that people may trip over tended to be of higher importance, while larger obstacles that people may bump into tended to have lower importance (P1, P3--P6).
Beyond hazards, risk indicator objects also had different importance levels based on their associated risk severity (P1--P6). For example, the stove indicator lights had higher importance as they indicated severe burning risks (P1). Similarly, crosswalks, pedestrian signals, tactile domes on curb cuts, traffic signs, and construction cones were rated as high-importance as they all indicated highly risky mobility tasks (e.g., crossing a street or tripping over construction obstacles; P1--P3, P5, P6). In contrast, sidewalks were considered less important due to their milder consequences (P1, P2, P4, P5). 

\textbf{Visual Difficulty.}
Four participants (P1, P3, P5, P6) considered visual difficulty a factor that affects object importance, with objects harder to perceive (e.g., low contrast, transparent) being more important to augment (P3, P5). 

\subsubsection{Design Guidelines for Object Augmentation}
\label{subsubsec:design_guidelines}
We summarize participants' preferences for augmenting important objects and distinguishing importance levels via three guidelines (DG 1--3).

\textbf{DG 1: Only use dynamic augmentations for crucial objects.}
Participants considered all four designs effective, including \textit{static outline} (P1--P6), \textit{solid overlay} (P1--P3, P6), \textit{flashing outline} (P1, P3), and \textit{text label} (P1, P2, P5). However, when augmenting multiple objects in a complex scene, all participants found the flashing augmentation visually distracting. They suggested that such dynamic augmentations should only be used for highly crucial and risky objects that require immediate attention, such as an approaching vehicle (P3) or construction cones (P5).

\textbf{DG 2: Balance between visibility and visual overload}.
Four participants (P1, P2, P4, P6) considered the solid overlay visually obvious but also overwhelming when augmenting too many objects (P1--P3, P5). They preferred the static outline for less visual blockage.
Similarly, 
four participants (P2--P5) considered the text label overly complex if rendered on multiple objects. P6 thus suggested augmenting objects with icon labels to reduce visual clutter.

\textbf{DG 3: Differentiate object importance on multiple visual dimensions.}
When augmenting important objects, five participants (P1--P5) considered it necessary to distinguish objects based on their importance levels.
Participants praised all four distinction methods and 
three participants (P1, P2, P5) suggested combining multiple designs (e.g., color and visual effect) to better distinguish object importance. For example, P1 and P5 suggested augmenting the most important objects using flashing outlines in a bright color (e.g., yellow) and less important ones using static outlines in a darker color (e.g., blue) for stronger visual contrast. 
\section{\system{}}
To demonstrate the concept of AR distinction for complex scenes, we designed and developed \system{}, a wearable AR system that detects and visually distinguishes objects of different importance levels through AR augmentations. 
We describe its design and implementation below.


\subsection{Augmentation Design}
\label{subsec:augmentation_design}

Following the design guidelines from Section \ref{subsubsec:design_guidelines} and one pilot study, \system{} supports three base augmentation designs and three AR distinction methods to distinguish objects of different importance. 
We detail the designs below.

\begin{figure*}[t]
    \centering
    \includegraphics[width=\textwidth]{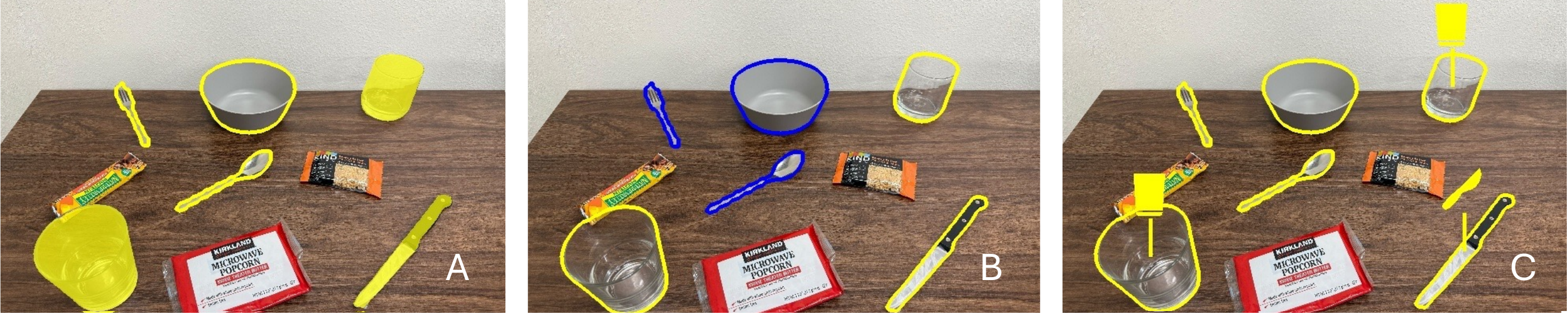}
    
  \caption{An illustration of the three AR distinction methods in \system{}. (A) \textit{By Form}: e.g., solid overlay vs. static outline, (B) \textit{By Color}: e.g., yellow vs. blue, (C) \textit{By Additional Visual Information}: e.g., static outline \& icon labels vs. static outline only. 
 }
  \Description{This figure illustrates the three AR distinction methods in \system{}. (A) Distinguishing object importance by form, with solid overlay on more important objects and outline on less important objects. (B) By Color, with yellow outline on more important objects and blue outline on less important objects. (C) By Additional Visual Information, with outline + icon labels for more important objects and only outline for less important objects. 
  }
  \label{fig:design}
\end{figure*}

\subsubsection{Base Augmentations}
\label{subsubsec:design}
We refined the four initial base augmentation designs following the design guidelines (DG 1--3).
Specifically, we replaced the text labels with icons to reduce visual complexity while maintaining necessary visual information (DG 2). While PLV liked the flashing effect for highly important objects (DG 1), we removed this design because it became indistinguishable from the static outline when object detection was unstable (the outline flickered due to occasional mis-detection). As a result, \system{} offers three base augmentations for important objects: (1) \textit{\textbf{Static Outline}}, a bright contour around the object to enhance its visibility and minimize visual blockage; (2) \textit{\textbf{Solid Overlay}}, a semi-transparent colored overlay over the object matching its exact shape; 
and (3) \textit{\textbf{Icon Label}}, a floating icon of the object and a line connecting the icon to the object center. The icons are selected to be visually recognizable and easily understandable. 


\subsubsection{AR Distinction Methods}
\label{subsubsec:differentiation}
Based on the factors (i.e., risk severity, visual difficulty) influencing object importance (Section \ref{subsubsec:formative_priority}), \system{} categorized objects into three categories: (1) \textit{primary-important} objects requiring immediate attention, (2) \textit{secondary-important} objects needing subtle augmentations, and (3) \textit{non-important} objects needing no augmentations.
Following the pilot study, we removed the \textit{By Visual Effect} method along with the flashing outline augmentation, ending with three distinction methods: 
(1) \textit{\textbf{By Form}}, which assigns different base augmentations to different importance levels (e.g., \textit{Solid Overlay} for primary-important objects vs. \textit{Static Outline} for secondary ones; Figure \ref{fig:design}A);
(2) \textit{\textbf{By Color}}, which assigns different colors when applying the same base augmentation to different importance levels (e.g., bright yellow outline for primary-important objects vs. darker blue for secondary ones; Figure \ref{fig:design}B); 
(3) \textit{\textbf{By Additional Visual Information}}, which adds additional information to primary-important objects (e.g., primary-important objects are augmented with both \textit{Static Outline} and \textit{Icon Label} vs. secondary ones with \textit{Static Outline} only; Figure \ref{fig:design}C). 

\subsubsection{Customization \& Combination}
\label{subsubsec:customization} 
To accommodate users' different preferences and visual needs (DG 3), \system{} allows flexible customization and combination of the AR distinction methods.
Color and opacity can be customized for all designs. Following prior research on PLV's perception of commercial AR devices \cite{zhao2017understanding}, we used yellow as the default color for its high visibility and offered green, blue, red, white, and cyan as options. Users can also adjust the thickness of \textit{Static Outline} (1 to 10 pixels) and the size of \textit{Icon Label} (from $50\times50$ to $240\times240$ pixels in 10-pixel steps).

Users can also combine multiple AR distinction methods to further enhance the visual differences between objects. For example, they can combine the \textit{By Form} and \textit{By Color} methods to augment primary-important objects with a yellow solid overlay and secondary ones with a blue outline.

\subsection{Prototype Implementation}
We prototyped \system{} by fine-tuning object recognition models for important objects and implementing the system on Microsoft HoloLens. We describe each system component below.

\subsubsection{Detection of Important Objects: Dataset Refinement and Model Finetuning} 
\label{subsec:model}
Based on the formative study (Section \ref{subsec:formative_findings}), we determined a list of important objects and their importance levels (primary- vs. secondary-important) for the kitchen and street navigation scenarios to enable object detection.
In the kitchen scenario, we included 11 object classes: three \textit{safety-related} classes (\textit{carafe, knife, scissors}), two \textit{visually challenging} classes (\textit{glasses, jars}), and six \textit{frequently used} ones (\textit{bowl, cup, fork, ladle, spatula, spoon}). Among them, we considered \textit{knife, scissors, carafe, glasses, jars} as primary-important and the rest as secondary-important.

In the outdoor scenario, we covered 21 object classes. All objects were considered \textit{safety-related} (\textit{curb, curb cut, fence, crosswalk, pedestrian, bench, sewer drain, fire hydrant, utility box, mailbox, lamp pole, construction cone, pedestrian signal, trash can, bicycle, motorcycle, railway track, sidewalk, vehicle, traffic sign, vehicle signal}). Meanwhile, one also belonged to the \textit{visually challenging} category (\textit{pedestrian signal}) and two belonged to the \textit{frequently used} category (\textit{traffic sign, vehicle signal}). We excluded most visually challenging objects (e.g., small objects, low-contrast objects, low-lighting objects) as they were based on low-level visual features instead of high-level objects. Among these objects, we considered \textit{curb, curb cut, crosswalk, sewer drain, construction cone, pedestrian signal, traffic sign, bicycle, motorcycle, railway track, and vehicle} as primary-important and the rest as secondary-important.
Based on the important object list, we constructed datasets to enable detection. For the kitchen scenario, we built the \textit{Kitchen-Importance} dataset of 20,052 images by combining and relabeling MS-COCO \cite{lin2015mscoco} and the Kitchen Affordance dataset \cite{lee2024cookar}. For outdoors, we constructed a dataset of 20,000 images, named \textit{Street-Importance}, by relabeling the Mapillary Vistas dataset \cite{neuhold2017mapillary}.

We fine-tuned an object segmentation model on the two datasets to detect important objects in each scenario. 
We selected the \textit{RTMDet} model \cite{lyu2022rtmdet} (the \textit{RTMDet-Ins-l} variant) due to its balance between recognition accuracy (mAP=0.437 on MS-COCO) and inference speed (33 FPS on an NVIDIA 4070 GPU). We initialized the model with pre-trained weights on MS-COCO and fine-tuned the model by freezing the backbone and only training the recognition head.
We refer to the fine-tuned models as \textit{RTMDet-Ins-l-Kitchen} and \textit{RTMDet-Ins-l-Street}, respectively, and evaluate them in Section \ref{subsec:recognition_accuracy}.

\subsubsection{System Framework}
\label{subsec:pipeline}
\system{} consists of a frontend on Microsoft HoloLens that senses the surrounding environment and renders object augmentations, and a backend on a remote server that receives the real-time video stream from HoloLens and recognizes important objects using the fine-tuned models. The video was streamed from HoloLens to the server using the \textit{hl2ss} library \cite{dibene2022hololens}, modified to support transmission via User Datagram Protocol (UDP). The recognition results and computed AR augmentation were serialized using \textit{Protocol Buffer} \cite{protobuf}, a cross-language protocol that supports compact data compression, and streamed to the frontend via UDP for augmentation rendering.

To accurately position the augmentations in 3D space, \system{} leveraged the RGB video input for 2D object localization and the environmental mesh captured by HoloLens for depth estimation. For each recognized object, 
we converted its 2D coordinates in image space into HoloLens' camera space using camera intrinsics from HoloLens' Research Mode API \cite{ungureanu2020hololens2researchmode}. 
We then cast a ray from the camera position toward the object center and rendered the augmentation at the point where the ray collided with the environmental mesh. Augmentations were scaled by distance to match the actual object size.
We rendered augmentations using the RawImage API in Unity 2022.3.25f1 \cite{unity3dUnityManual}.

\begin{figure*}[h!]
    \centering
    \includegraphics[width=\textwidth]{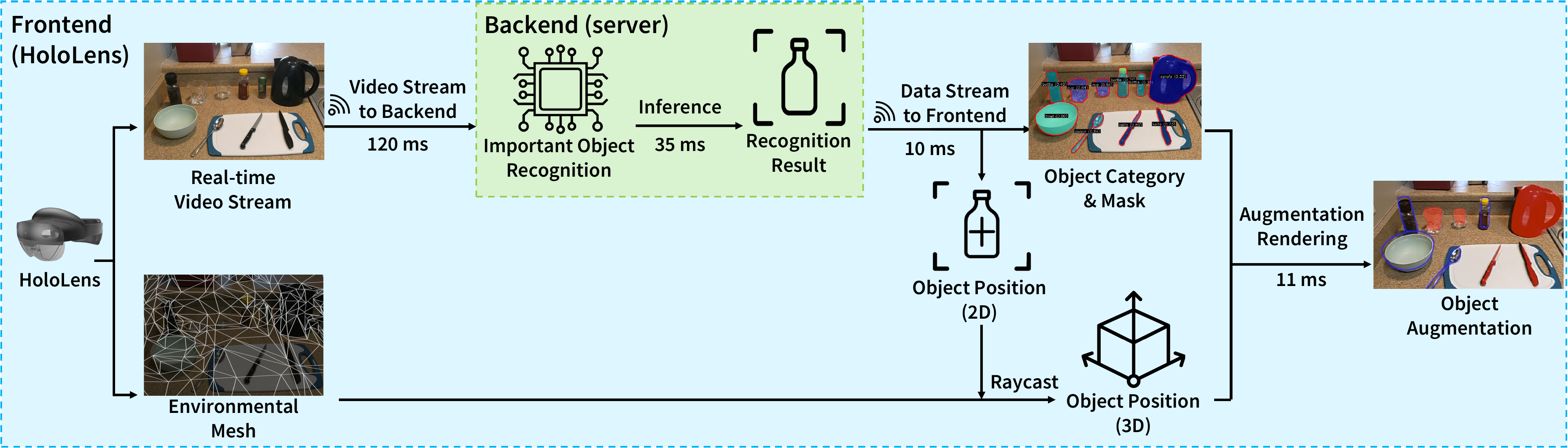}
  \caption{System pipeline of \system{}: the HoloLens (frontend) streams video to the backend; the backend runs the fine-tuned models to recognize and segment important objects and streams the result to the HoloLens; the HoloLens receives each object's 2D position, raycasts it onto the 3D environmental mesh to locate the 3D position, and renders augmentations. We label each stage's latency in the figure. \colorchange{The overall latency is 176 ms and the frame rate is 28.56 FPS.}}
  \Description{This graph shows the system pipeline of SceneGlance: the HoloLens (i.e., frontend) streams video to the backend; the backend runs the fine-tuned models to recognize and segment important objects; the backend streams the recognition result to the HoloLens; the HoloLens locates the 3D position of objects by raycasting the 2D location of each object onto the 3D environmental mesh and renders augmentations. The overall latency is 176 ms and the frame rate is 28.56 FPS.}
  \label{fig:pipeline}
\end{figure*}
\subsection{Technical Evaluation of the System}
\label{sec:technical_evaluation}
We performed a technical evaluation of \system{} on two aspects: real-time performance and recognition accuracy.

\subsubsection{Real-Time Performance}
\label{subsec:framerate_delay}
We ran \system{} continuously for three minutes and measured the frame rate and mean latency of each stage in the pipeline (Section \ref{subsec:pipeline}).
Specifically, video streaming from HoloLens to the backend ran at 29.78 FPS with a 120 ms latency; model inference and post-processing ran at 28.56 FPS with a 35 ms latency; result streaming from the backend to HoloLens ran at 28.56 FPS (as it was capped by model inference) with a 10 ms latency; and object localization and rendering ran at 28.56 FPS with an 11 ms latency. Overall, \system{} ran at 28.56 FPS with a 176 ms per-frame latency, reaching near real-time performance. We offset the latency by reprojecting augmentations based on the user's pose at the time of capture, so that the perceived visual delay of static object augmentations was 11 ms; however, this technique does not apply to moving objects (e.g., pedestrian in outdoor navigation). 

\begin{figure*}[t]
    \centering
    \includegraphics[width=\textwidth]{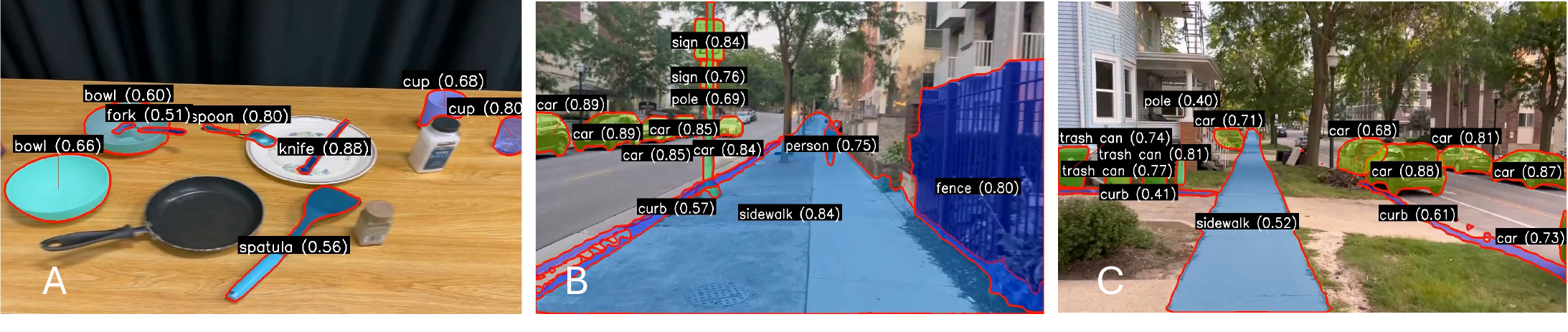}
  \caption{Example inference results of the two fine-tuned models on test images in the kitchen (A) and outdoor environment (B, C). These examples illustrated the models' robustness in complex environments against partial occlusion. }
  \Description{This figure shows some example inference results of the two fine-tuned models on test images in the kitchen and outdoor environment. These examples illustrated the models' robustness in complex environments against partial occlusion.}
  \label{fig:model_demo}
\end{figure*}

\subsubsection{Recognition Accuracy}
\label{subsec:recognition_accuracy}
We evaluated the recognition accuracy of the two fine-tuned models described in Section \ref{subsec:model} on the test subsets of our kitchen and street object datasets. For each dataset, we used a random train/validation/test split of 80\%/10\%/10\%. We used three common metrics \cite{padilla2021comparative}: average precision at a 50\% Intersection over Union (IoU) threshold (AP@50), at a 75\% IoU threshold (AP@75), and mean average precision (mAP), defined as the mean of average precision across IoU thresholds
from 50\% to 95\% with step 5\%. These IoU-based metrics holistically assess how accurately the predicted masks align with the ground truth \cite{everingham2010pascal}. 


We compared the two fine-tuned models against the baseline model, the RTMDet-Ins-l model pretrained on \textit{MS-COCO} without fine-tuning.
The results (Table \ref{tab:recognition_accuracy}) showed that our fine-tuned models outperformed the baseline model across all three metrics, reaching mAPs of 0.435 (Kitchen) and 0.324 (Street). Figure \ref{fig:model_demo} shows examples of the fine-tuned models performing on test images. Our models demonstrated strong robustness in complex scenes, maintaining high accuracy in busy kitchen and street environments from an egocentric perspective. The models were also robust against partial occlusion (e.g., a spoon in a bowl).


\begin{table}[htbp]
\centering
\small
\begin{tabular}{p{2.70cm}|p{0.4cm}p{0.72cm}p{0.72cm}}
\toprule
  \textbf{Model} & \textbf{mAP} & \textbf{AP@50} & \textbf{AP@75} \\
\hline
RTMDet-Ins-l (baseline) & 0.190 & 0.313 & 0.200\\
RTMDet-Ins-l-Kitchen & \textbf{0.435} & \textbf{0.683} & \textbf{0.477} \\
\hline
RTMDet-Ins-l (baseline) & 0.133 & 0.221 & 0.133 \\
RTMDet-Ins-l-Street & \textbf{0.324} & \textbf{0.558} & \textbf{0.337}\\

\bottomrule
\end{tabular}
\caption{Recognition accuracy of the fine-tuned and baseline models. Both fine-tuned models outperformed the baseline.}
\Description{This table shows the recognition accuracy of the fine-tuned and baseline models. Both fine-tuned models significantly outperformed the baseline RTMDet-Ins-l model and achieved high mAP.}
\label{tab:recognition_accuracy}
\vspace{-5ex}
\end{table}

\subsubsection{False Negatives and False Discoveries}
\colorchange{
We analyzed the false negative rate (indicating the rate of unaugmented important objects) and the false discovery rate (indicating the rate of wrongly augmented non-important objects) \cite{padilla2021comparative} of the two fine-tuned models. 
Across object categories, the RTMDet-Ins-l-Kitchen model has a false negative rate of 29.3\% and a false discovery rate of 29.7\%; the RTMDet-Ins-l-Street model has a false negative rate of 42.3\% and a false discovery rate of 32.2\%. We report per-category rates in Tables~\ref{tab:kitchen_recognition} and \ref{tab:outdoor_accuracy} in Appendix~\ref{app:recognition_accuracy}.
}
 
\colorchange{In both scenarios, the two error types were associated with different visual characteristics.
In the kitchen scenario, false negatives were most common for transparent objects, such as glasses (39.4\%) and jars (37.7\%), echoing the visually difficult objects identified in the formative study (Section~\ref{subsec:formative_findings}). False discoveries were most common for objects that are visually similar to other kitchen items, such as ladles (39.4\%), bowls (36.5\%), and scissors (34.9\%).
In the outdoor scenario, false negatives were most common for surfaces that lack distinct visual features, such as curb cuts (89.0\%) and crosswalks (79.1\%), and for objects that vary widely in visual appearances, such as fences (73.1\%). False discoveries were most common for objects that are visually similar to other common obstacles in the scene, such as sewer drains (51.1\%), utility boxes (42.6\%), and benches (39.5\%).}
\section{Study I: Using \system{} in a Complex Mock-up Kitchen}
\label{sec:kitchen_study}
We first evaluated PLV's complex scene perception with and without \system{} in a kitchen environment, understanding how AR distinction designs change people's viewing strategy and mental map development. We also used \system{} as a technology probe to uncover the unique perception challenges, strategies, and preferences of PLV viewing AR-enhanced complex scenes, shedding light on practical AR system designs for complex reality. 


\subsection{Methods}
\subsubsection{Participants}
\label{subsubsec:kitchen_participants}
We recruited 12 participants with low vision (R1-R12) through contact lists from local low vision clinics and communities. A participant was eligible for the study if they were at least 18 years old and had best-corrected visual acuity between 20/100 and 20/800 in the better eye.
Table \ref{tab:demographics_kitchen} in Appendix~\ref{app:demographics_kitchen} details participants' demographic information. Participants were aged between 19 and 73 ($M=48.75$, $SD=18.68$), including six males and six females. All participants except R2, R4, and R10 were legally blind.
R1 and R3 experienced peripheral vision loss, and six participants (R2, R4, R7, R10--R12) experienced central vision loss. 
Six participants (R1, R4, R9--R12) had prior experience with AR, but none used it regularly.
Participants were compensated \$25 per hour. The study was approved by our university IRB.

\subsubsection{Apparatus}
We set up a mock-up kitchen countertop with a size of $150~cm \times 60~cm$ in a well-lit lab space. 
To simulate a complex kitchen environment, we prepared 12 categories of kitchen objects, including four primary-important (knives, scissors, glasses, jars), three secondary-important (bowls, spoons, forks), and five non-important (snacks, fruits, cutting board, sponge, dish cloth) following the formative study.

\begin{figure*}[tbp]
    \centering
    \includegraphics[width=\textwidth]{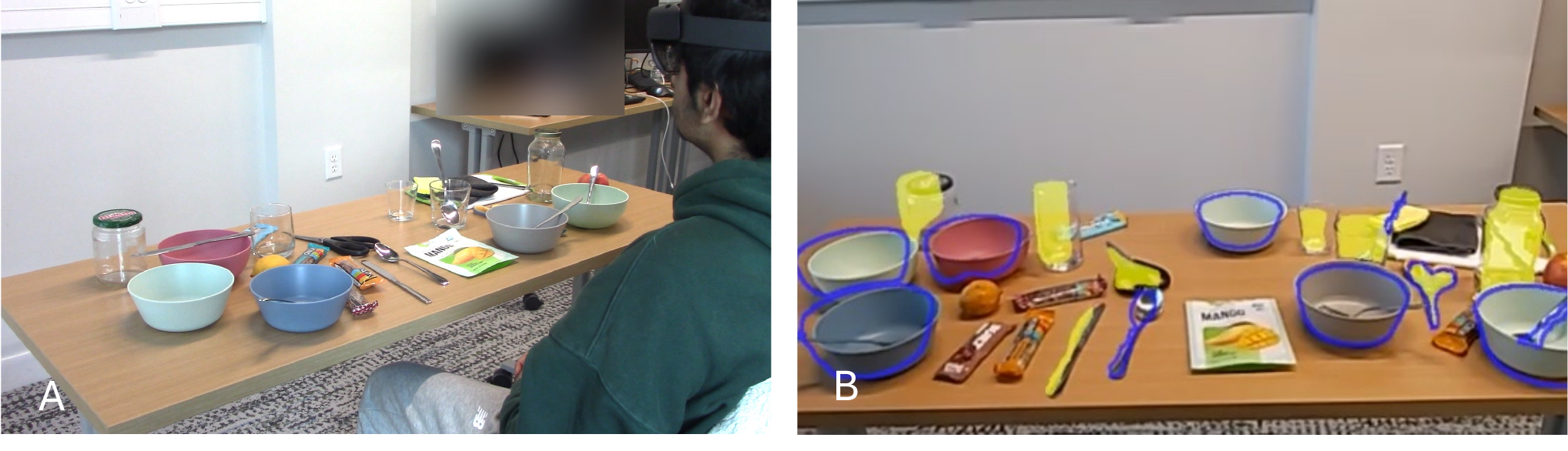}
  \caption{Kitchen Countertop Perception Task. (A) Participants sat at a mock-up kitchen table with 30--33 kitchen objects, contextualized in the scenario that they were going to cook on this table and needed to familiarize themselves with the table sufficiently for cooking. (B) Participants' view in \system{} where primary-important objects (glasses, jars, knives, scissors) were augmented in semi-transparent yellow solid overlay, and secondary-important objects (bowls, spoons, forks) augmented in blue outline. Non-important objects (snacks, cutting board, sponge, dish cloth, fruits) were not augmented.}
  \Description{Demonstration of the kitchen countertop perception task task. (A) Participants sat at a mock-up kitchen table with 30--33 kitchen objects. Participants were contextualized in the scenario that they were going to cook on this table and needed to familiarize themselves with the scene sufficiently enough for cooking. (B) An illustration of participants' view in \system{} where primary-important objects (glasses, jars, knives, scissors) were augmented in semi-transparent yellow solid overlay, and secondary-important objects (bowls, spoons, forks) augmented in blue outline. Non-important objects (snacks, cutting board, sponge, dish cloth, fruits) were not augmented.}
  \label{fig:summative_indoor_demo}
\end{figure*}

With these kitchen objects, we generated six table layouts with similar amount and variability of objects to achieve similar scene complexity for the six trials (details in Section \ref{sec:procedure}). Each layout contained 30--33 objects, with 10 or 11 objects from each object importance category (i.e., primary-important, secondary-important, and non-important). The objects were randomly placed on the table, and utensils (i.e., forks, spoons, knives) were either placed on the table or in containers (i.e., bowls, glasses). Figure \ref{fig:summative_indoor_demo}A shows an example of the table layout and Figure \ref{fig:summative_indoor_demo}B shows a participant's view of the kitchen countertop with objects augmented by \system{}. 


\subsubsection{Procedure} \label{sec:procedure}
The study consisted of one session lasting three hours,
where participants conducted scene perception tasks in three conditions: (1) \textit{Reality baseline}: not wearing an AR device but using their own vision with best correction (e.g., eyeglasses); (2) \textit{AR baseline}: wearing the AR glasses with all important objects augmented by the same base design; and (3) \textit{\system{}}: wearing \system{} that distinguished primary- and secondary-important objects with AR distinction designs. We describe the study details below.

\textit{\textbf{System Tutorial and Customization}}. We started by collecting participants' demographics, visual conditions, and prior AR experience. Then, we prepared a tutorial scene with 31 objects across all 12 categories 
randomly placed on the table, and demonstrated \system{}.
We first walked through all three base augmentations. For each design, participants freely customized the augmentation parameters (e.g., color, opacity, outline thickness, icon label size) and discussed what they liked and disliked about the design. Participants then chose their preferred augmentation if only one type of augmentation could be used on both primary- and secondary-important objects (for the AR baseline).
Next, we demonstrated the three AR distinction designs and asked about participants' feedback and preferences.
Participants then freely explored the tutorial scene using \system{} with their preferred designs until they felt fully familiar with the system and its augmentations.

\textit{\textbf{Scene Perception Tasks}}.
Participants then completed scene perception tasks on the kitchen countertop. We contextualized them in a scenario where they would cook on the kitchen countertop and needed to familiarize themselves with the table sufficiently. They could proceed at their own pace and walk around the table as needed. Once participants felt sufficiently familiar with the scene, they verbally described it without looking at the table, and drew a mental map of all objects they remembered. We also asked which objects caught their attention first and why. 

Participants completed the tasks under three conditions: the Reality baseline, the AR baseline, and \system{}, with two trials each, totaling six. 
After the two trials in each condition, participants reflected on their experience and perceived effectiveness of scene perception under that condition. 
We counterbalanced the order of the conditions using Latin Square and randomly mapped the six table layouts to the six trials for each participant. 

\textit{\textbf{Exit Interview}}. We concluded the study with an interview where participants compared their scene perception experience across the three conditions. Participants also discussed whether two importance levels were sufficient and brainstormed augmentation designs to better distinguish object importance.

\subsection{Data Analysis}
\label{subsec:kitchen_data_analysis}
We collected both quantitative and qualitative data from the study and analyzed them as follows.

\subsubsection{Quantitative Analysis}
We evaluated participants' mental maps with two metrics: (1) \textit{Object Recall}, defined as the proportion of correctly recalled objects among all objects present on the table, to reflect how well participants perceived the scene under each condition; (2) \textit{Recall Ratio}, defined as the proportion of correctly recalled objects at each importance level among all correctly recalled objects, to reflect how participants allocated their attention across importance levels.
We computed \textit{Object Recall} at five levels: overall, all important objects (i.e., primary- and secondary-important combined), primary-important, secondary-important, and non-important objects. We computed \textit{Recall Ratio} at four levels: all important, primary-important, secondary-important, and non-important objects. We had nine measures in total.

For all measures, we had one within-subject factor, \textit{Condition}, with three levels: \textit{Reality baseline}, \textit{AR baseline}, and \textit{\system{}}. To validate the counterbalancing, we included a between-subjects factor \textit{Order}, \colorchange{and found no significant effect of \textit{Order} on any measures}. We applied the Shapiro-Wilk normality test \cite{shapiro} to all nine measures. Five measures (overall recall, important object recall, primary-important recall, secondary-important recall, and secondary-important ratio) were normally distributed, and four measures (non-important recall, important ratio, primary-important ratio, and non-important ratio) were not. If a measure was normally distributed, we fitted a Linear Mixed-Effects (LME) model and computed the ANOVA table to evaluate the effect of \textit{Condition} \cite{maxwell2017designing, kuznetsova2017lmertest, wang2024gazeprompt}, and used Tukey's HSD for \textit{post-hoc} comparison if significance was found \cite{tukey1949comparing}. For non-normal measures, we used Aligned Rank Transform (ART) ANOVA \cite{2011ART} to test significance and ART-C \cite{2021ARTC} for \textit{post-hoc} comparison. \colorchange{Since we ran tests on nine measures, we applied Bonferroni correction to the significance threshold ($\alpha=0.05/9=0.0056$) \cite{dunn1961multiple, armstrong2014use}.} We calculated the effect size using partial eta squared ($\eta_p^2$), with 0.01, 0.06, and 0.14 as the thresholds of small, medium, and large effects, respectively \cite{cohen2013statistical}.


Moreover, we categorized the first object noticed in each trial by importance level and summarized the distribution across the three conditions in Table \ref{tab:first_noticed}. We collected 24 valid responses for each of the Reality baseline and \system{} conditions. In the AR baseline, two participants (R1, R6) each reported in one trial that no object stood out at first glance as they felt confused by the augmentations, resulting in 22 valid responses. To test whether the distribution of first-noticed objects differed across conditions, we applied the Pearson chi-square test \cite{pearson1900x} followed by pairwise chi-square tests with Bonferroni correction \cite{dunn1961multiple}.


\subsubsection{Qualitative Analysis}
\label{subsubsec:kitchen_qualitative_analysis}
We video-recorded all studies. 
All recordings were transcribed using an automatic transcription service and manually revised by the research team. We analyzed the transcripts using thematic analysis \cite{braun2006using, braun2024thematic}. Two researchers independently open-coded three participant transcripts (25\% of the data) to develop an initial codebook upon agreement. Then, one researcher coded the remaining data based on the codebook. When a new code emerged, researchers discussed it and added it to the codebook upon agreement. 
Codes were grouped into themes and sub-themes guided by our research focus, including perception challenges of complex scenes for PLV, effects of augmentation and distinction on attention and scene understanding, and design challenges of AR augmentations in complex scenes.

\subsection{Findings} 

We report findings on how PLV perceived the complex scene under the three conditions, revealing their viewing strategy and shifts in attention allocation with and without \system{}. \colorchange{We found that \system{} shifted attention toward primary-important objects but reduced overall scene recall, revealing an attention-recall tradeoff in augmenting complex scenes.} 
Our study also revealed unique challenges of applying AR augmentations to complex, cluttered scenes, providing critical design insights on \textit{what objects to augment and distinguish} and \textit{how to effectively distinguish them} for future AR systems. We detail our findings below. 


\subsubsection{Visual Strategies for Perceiving Complex Scenes}
\label{subsubsec:bare_eye_strategy}
The crowded kitchen countertop with over 30 objects posed significant perceptual challenges. Without any augmentations, participants adopted several strategies to perceive and organize the scenes.

\textbf{Linear scanning \& grouping.}
Most participants (8/12; e.g., R1, R7) scanned the scene linearly from one end to the other, sometimes repeating the scan multiple times. As R7 described: ``With bare eyes, I was kind of just going left to right multiple times in a row to get a good scan of the whole table.'' During linear scanning, four participants (R2, R8, R11, R12) grouped nearby objects of the same category and memorized object groups instead of each individual object. As R2 described: ``Everything was very compartmentalized. Four glasses on the right side, three bowls on the back, one in the center left-ish.''

\textbf{Anchor-based mental map construction.}
Four participants (R4, R6, R9, R11) actively looked for certain objects as ``anchors'' and developed mental maps for the rest of the scene relative to them. As R9 explained: ``I've kind of been using [bowls, cups, and jars] as anchors for remembering the table [...] in my head I just made a map: okay, the bowls are here, the cups are here, the jars there. And then I try to remember what's in them.''
Participants chose anchor objects with certain characteristics: R6 and R9 selected large objects (e.g., bowls, a large pack of snacks) as they were visually salient, R11 chose objects that were unique in the scene (e.g., the only cutting board) because ``there aren't multiples of it, so it's easier to remember exactly where those items are,'' and R4 focused on breakable objects (e.g., glasses, jars) as he was concerned about knocking them over. This anchor-based mental model provides insights into what should be augmented in complex scenes. We discuss this further in Section \ref{subsec:design_implications}. 

\textbf{Relating to everyday activities.}
Three participants (R5, R10, R12) interpreted the scene by relating it to everyday activities, such as a table setting for a meal (R5, R12) or a kitchen around mealtime (R10). As R10 described: ``I just see bowls, almost like someone's ready to eat or they just finished eating. I see the cutting board, it looks like it had some action with the scissors and the knife.'' R12 further said that this kind of association guided his attention and mental model. For example, he focused on a bowl with a utensil on the table since he imagined it as part of a meal: ``It almost felt like I was at a restaurant or at home, and it could [look like] food [...] right in the middle of the table and a utensil in it. Some of those patterns or schemas are easy to recognize and immediately where some of my attention goes.''

Despite these strategies, it remained mentally demanding for some participants (R9--R12) to perceive the complex scenes because they lacked visual guidance on which objects to pay attention to first. As R9 explained: ``I feel like [without augmentations] I was looking at more stuff. I didn't have a clear priority of what to look at, you know? [With \system{}] it was like: okay, I remembered the utensils first, and I felt like that was what I needed to remember, whereas here I'm looking at everything all at once.''


\begin{table*}[htbp]
\centering
\small
\begin{tabular}{p{2.8cm}p{2.8cm}p{2.8cm}p{2.8cm}p{1.8cm}}
\toprule
  \textbf{Condition (\# trials)} & \textbf{Primary-important} & \textbf{Secondary-important} & \textbf{Important (Primary + Secondary)} & \textbf{Non-important} \\
\hline
Reality baseline (24) & 5 (20.8\%) & 8 (33.3\%) & 13 (54.2\%) & 11 (45.8\%) \\
AR baseline (22) & 10 (45.5\%) & 7 (31.8\%) & 17 (77.3\%) & 5 (22.7\%) \\
\system{} (24) & 19 (79.2\%) & 2 (8.3\%) & 21 (87.5\%) & 3 (12.5\%) \\
\bottomrule
\end{tabular}
\caption{Distribution of first-noticed objects across the three conditions. Each participant reported their first-noticed objects in two trials per condition. In the AR baseline, two participants (R1, R6) each reported in one trial that nothing stood out at first glance, resulting in 22 valid responses.}
\Description{This table shows the distribution of first-noticed objects across the three conditions. Participants increasingly noticed primary-important objects first as augmentation and distinction were introduced, from 20.8\% in the Reality baseline to 45.5\% in the AR baseline to 79.2\% with \system{}.}
\label{tab:first_noticed}
\vspace{-5ex}
\end{table*}

\subsubsection{AR Importance Distinction Reshaped Attention Allocation}
\label{subsubsec:attention_allocation}
We report how \system{} affected attention allocation in the complex scene. Compared to the Reality baseline, 
\system{} effectively shifted participants' attention toward primary-important objects, reflected in both object recall ratio and first-noticed object distribution. 
We also summarize qualitative feedback on how the AR distinction supported PLV's perception of complex scenes.

\textbf{Attention allocation across conditions.}
We found a significant effect of \textit{Condition} on the recall ratio of primary-important objects ($F_{2,58}=6.11$, $p=0.004$, $\eta_p^2=0.17$). Both the AR baseline ($diff=13.75$, $p=0.018$) and \system{} ($diff=15.56$, $p=0.006$) showed a significantly higher recall ratio than the Reality baseline. No significant difference was found between \system{} and the AR baseline ($p=0.927$). No significant effect was found for the recall ratio of \colorchange{all important objects ($F_{2,58}=3.60$, $p=0.034 > 0.0056$ with correction) or} secondary-important objects ($F_{2,58}=0.31$, $p=0.737$). 

We also found a significant effect of \textit{Condition} on the distribution of first-noticed objects ($\chi^2(4)=17.45$, $p=0.002$). Pairwise comparisons showed that \system{} significantly differed from the Reality baseline ($p<0.001$), with participants first noticing primary-important objects in 79.2\% of the trials with \system{} in contrast to 20.8\% in the Reality baseline (Table \ref{tab:first_noticed}). The AR baseline fell in-between with no significant difference from \system{} ($p=0.149$) or the Reality baseline ($p=0.426$). 
These results indicated that
\system{} effectively shifted participants' attention toward \colorchange{primary-important objects that they most} desired to see.


\textbf{Visual strategy changes.}
Participants reported that the importance distinction of \system{} helped them better perceive the complex scenes compared to the AR baseline. Augmenting all important objects equally did not fully address the high mental demand of perceiving complex scenes. Six participants (e.g., R1, R9) reported that, with over 20 objects augmented in the same way in the AR baseline, they still felt confused about where to look first. As R1 said: ``The fact that everything was highlighted in the same color, so there's highlights over here, there's highlights there [...] where do I look first?''
In contrast, the importance distinction from \system{} enabled participants to adopt new visual strategies to address such disorientation issues. We describe their new visual strategies below.


\textit{\underline{Mental snapshot from augmentation distribution.}}
Before perceiving individual objects, participants used the distribution of different augmentations to establish a rough impression of the scene (R4, R6, R10). As R4 reflected: ``[With the AR distinction] there's a lot of icons over here but not a lot of icons over here, [it] helps understand more of a general danger zone [...] spatially remembering where more things that are fragile versus not [...] having that differentiation, it's a nice snapshot when you assess the scene: okay, there's a high concentration of glass over here and over here, but in this space there's not as much.''

\textit{\underline{Hierarchical scanning with visual guidance.}}
Using more pronounced augmentations for more important objects provided visual guidance for participants to start from primary-important objects before moving to secondary- and non-important ones (10/12; e.g., R5, R8). 
Since the AR distinction divided the scene into smaller object groups by importance levels, participants could focus on one group at a time instead of processing all augmented objects at once (in the AR baseline), which was easier to manage mentally (6/12; e.g., R5, R7). 
As R5 said: ``Everything wasn't the same. There was a difference. You know, [with the AR distinction] I looked and tried to figure out [primary-important objects]: oh, these are jars, these are glasses, these are scissors, okay, did all that first. The next thing, okay, now what's the other highlighting. It's in blue. Let me check, oh, those are the bowls [...] So instead of overwhelming me with everything all [augmented equally] at once, it gave me a chance to sort it out.''



\begin{figure*}[h!]
    \centering
    \includegraphics[width=\textwidth]{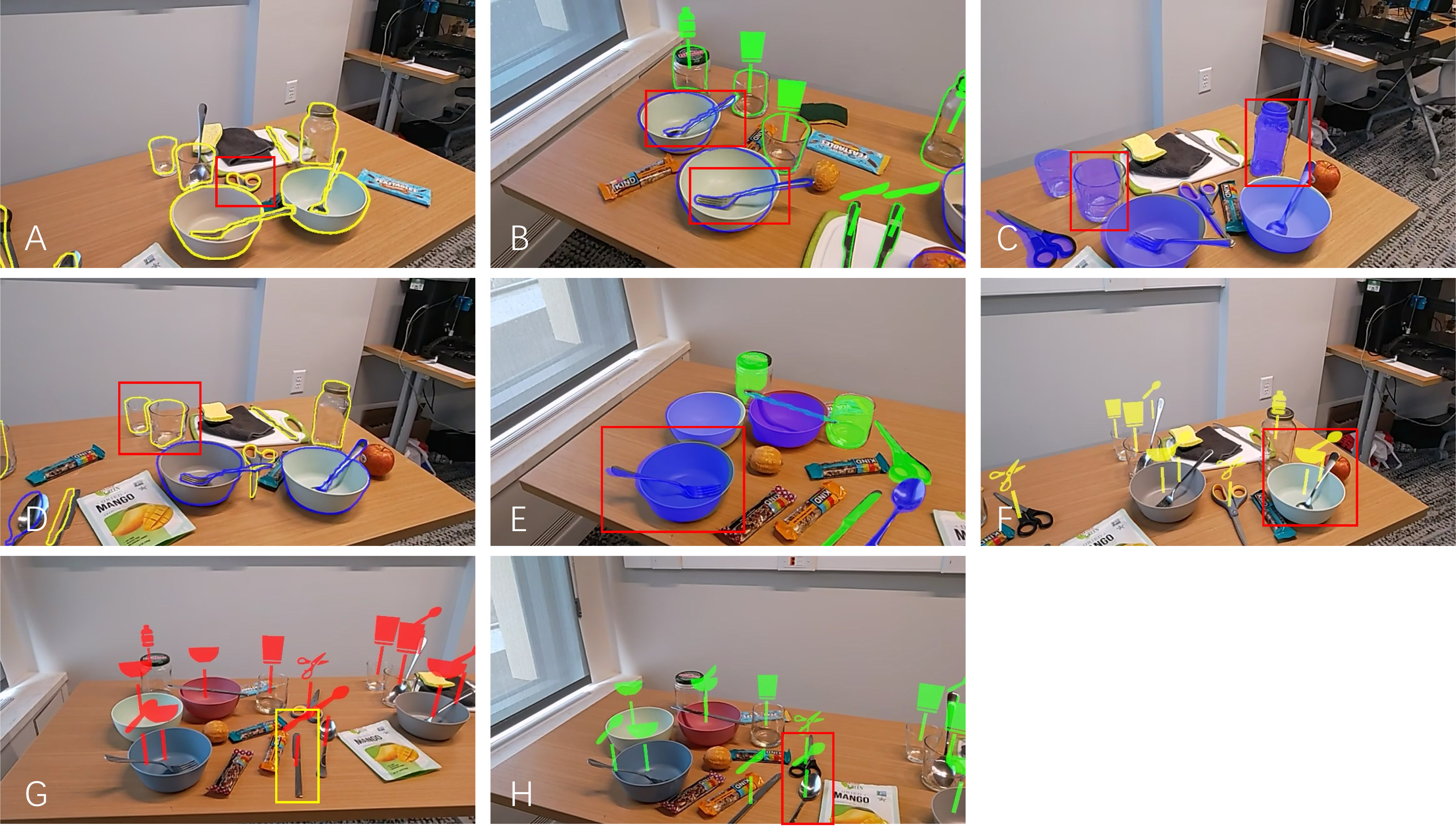}
  \caption{Examples of perception challenges in Study I. (A) \textit{Partial occlusion}: the outline of the scissors handle was misidentified as a ``heart,'' as the outline did not reveal the full object. (B) \textit{Visual similarity}: outlines of a fork and spoon looked similar because outlines did not preserve fine details such as fork tines. (C) \textit{Solid overlay hindering distinguishing similar objects}: the solid overlay covered the lid of a jar, making it indistinguishable from a glass. (D) \textit{Adjacent augmentations merging}: outlines of two adjacent glasses merged and appeared like a ``pitcher.'' (E) \textit{Solid overlays merging}: solid overlays of a bowl and a fork inside it merged into a single shape looking like ``a bowl with a handle.'' (F) \textit{Icon spatial misalignment}: icon labels of a bowl and a spoon inside it could not convey their spatial relationship (containment). (G) \textit{Icon label occluding the augmented object itself}: the connector line of the icon label of a knife occluded the upper part of the knife. (H) \textit{Icon label occluding objects behind}: the icon label of the spoon occluded part of the scissors behind it.}
  
  \Description{This figure contains eight sub-figures A--H, corresponding to eight example images for the perception challenges in Study I. (A) \textit{Partial occlusion}: a pair of scissors was partially occluded by a bowl, and only the handle was visible. The outline of the scissor handle was misidentified as a ``heart,'' as the outline did not reveal the full object. (B) \textit{Visual similarity}: outlines of a fork and spoon looked similar because outlines did not preserve fine details such as fork tines. (C) \textit{Solid overlay hindering distinguishing similar objects}: the solid overlay covered the lid of a jar, making it indistinguishable from a glass. (D) \textit{Adjacent augmentations merging}: outlines of two adjacent glasses merged and appeared like a ``pitcher.'' (E) \textit{Solid overlays merging}: solid overlays of a bowl and a fork inside it merged into a single shape looking like ``a bowl with a handle.'' (F) \textit{Icon spatial misalignment}: icon labels of a bowl and a spoon inside it could not precisely convey their spatial relationship (containment). (G) \textit{Icon labels occluding the augmented objects}: the connector line of the icon label of a knife occluded the upper part of the knife. (H) \textit{Icon labels occluding objects behind}: the icon label of the spoon occluded part of the scissors behind it.}
  \label{fig:indoor_problems}
\end{figure*}

\subsubsection{New Perception Challenges Posed by AR Augmentations}
\label{subsubsec:perceptual_challenges}

While \system{} shifted participants' attention toward primary-important objects (Section \ref{subsubsec:attention_allocation}), our study also revealed new barriers posed by AR augmentations to complex scene perception. We found a significant effect of \textit{Condition} on overall recall ($F_{2,58}=5.74$, $p=0.005$, $\eta_p^2=0.17$), where participants correctly recalled significantly fewer objects in both the AR baseline ($diff=-0.088$, $p=0.005$) and \system{} ($diff=-0.079$, $p=0.013$) than the Reality baseline, with no significant difference between \system{} and the AR baseline ($p=0.944$).
\colorchange{This decline was mainly driven by non-important objects, recalled significantly less under both AR conditions than the Reality baseline} ($F_{2,58}=6.76$, $p=0.002$, $\eta_p^2=0.19$).
No significant effect was found for important objects overall ($F_{2,58}=2.34$, $p=0.105$), primary-important objects ($F_{2,58}=0.51$, $p=0.605$), or \colorchange{secondary-important objects ($F_{2,58}=5.12$, $p=0.009 > 0.0056$ after correction).} 

\colorchange{These results suggest an \textit{attention-recall tradeoff} of the multi-object AR augmentations---shifted attention toward primary-important objects with reduced recall of non-important objects---}indicating potential challenges posed by AR augmentations in cluttered scenes. We found that some augmentations amplified existing visual difficulties, such as distinguishing partially occluded or visually similar objects, while augmenting multiple objects further introduced new challenges, such as adjacent augmentations clustering into confusing shapes and spatial misalignment of icon labels. We detail these challenges below.

\textbf{Amplifying confusion from partially occluded objects.}
One main challenge PLV faced was perceiving partially occluded objects, such as a utensil inside a bowl or scissors behind a bowl (8/12; e.g., R2, R9). In the cluttered kitchen, such occlusion was common, as nearby objects could hide distinguishing features (e.g., fork tines, scissor handles), making occluded objects hard to identify. For example, R2 misidentified the scissors as a knife as they were partially occluded by a bowl and only the blades were visible. 
On top of this challenge, the Static Outline and Solid Overlay only augmented the visible part, which could even amplify the confusion (8/12; e.g., R4, R10). For example, when perceiving the occluded scissors with the static outline, R8 misidentified the handle as ``a heart'' (Figure \ref{fig:indoor_problems}A).

The icon label helped participants identify partially occluded objects by visualizing the complete shape and indicating its category (5/12; e.g., R3, R8). As R8 said: ``[The icon] really makes the object clear as to what it is. It really defines that that is scissors, that is a knife. [Without augmentation] I would have had to move myself closer to the object to figure it out, versus the icon was like popped up so you knew [an object] was behind or around something.''


\textbf{Obscuring differences between visually similar objects.} 
Participants also struggled to distinguish similar objects with subtle differences (e.g., spoons and forks, or jars and glasses; 7/12; e.g., R2, R12). As with partial occlusion, icon labels helped because they conveyed object type through clearly distinguishable icons (R3, R4, R5, R11). However, the AR outline did not help as it did not preserve fine shape details (e.g., fork tines), providing similar outlines for different objects (R5, R9, R11, R12). For example, the spoon and fork in Figure \ref{fig:indoor_problems}B looked more similar due to similar AR outlines. The solid overlay further exacerbated this problem by obscuring object details, removing the subtle visual differences (e.g., shape, color) that participants relied on to distinguish similar objects (9/12; e.g., R4, R8). For example, R8 reported that the solid overlay made it harder to distinguish jars from glasses as it obscured whether the object had a lid: ``[with the overlay] I can't tell if there's a cover on [the glass]'' (Figure \ref{fig:indoor_problems}C).



\textbf{Adjacent augmentations blending in}.
We found that adjacent objects with similar importance levels posed unique challenges as the same augmentations clustered together and introduced visual confusion.  
The augmentations (e.g., outline, solid overlay) could overlap or merge, creating misleading shapes. For example, five participants (e.g., R1, R9) reported that multiple adjacent outlines would intersect to ``look like one big highlight'' (R9) and be mistaken for one object. R1 also misidentified the outlines of two adjacent glasses of different heights as a pitcher, with the shorter glass as the ``handle'' and taller one as the ``body'' (Figure \ref{fig:indoor_problems}D). Similarly, solid overlays on multiple adjacent objects would visually merge to create one shape that was hard to identify (6/12; e.g., R5, R12). For example, R12 misidentified a bowl with an extruding fork as ``a bowl with a handle'' from the two combined overlays (Figure \ref{fig:indoor_problems}E).
In contrast, icon labels did not suffer from this ``merging'' problem as they were spatially separated above the objects (R5, R10).

\textbf{Spatial misalignment of icon labels}.
Despite the advantages of icon labels mentioned above, they do not directly augment the original objects, generating spatial misalignment. This offset resulted in multiple problems in a complex scene:
First, as icons did not precisely indicate object boundaries, they failed to convey the exact relative position and relationships between overlapping objects (5/12; e.g., R2, R6). For example, when augmenting a spoon inside a bowl, R6 reported that both icons floated side-by-side at similar heights, giving no indication of whether the spoon was inside or beside the bowl (Figure \ref{fig:indoor_problems}F). 
Second, as icons floated above the augmented objects, they created new occlusion, blocking part of the augmented object itself (R2, Figure \ref{fig:indoor_problems}G), objects behind them (R5, R7, R9, Figure \ref{fig:indoor_problems}H), or the icons of other objects (R7). 

Moreover, the spatial misalignment required participants to consistently switch attention between icons and their corresponding objects, causing increased mental load (5/12; e.g., R1, R6). As R6 elaborated: ``I'd rather have the outline right on the bowl than a little [...] arrow pointing to it. It's kind of like: okay, I see a scissors [icon]. Now, where's the scissors? Oh, the scissors is down there.'' 
Some participants (R1, R5) also raised that in tasks requiring object interaction (e.g., cooking), the icons would mislead them as they ``tend to reach for the icon'' (R1).
\begin{table*}[htbp]
\centering
\small
\begin{tabular}{p{4cm}cccccccccccc}
\toprule
\textbf{AR Distinction Design} & \textbf{R1} & \textbf{R2} & \textbf{R3} & \textbf{R4} & \textbf{R5} & \textbf{R6} & \textbf{R7} & \textbf{R8} & \textbf{R9} & \textbf{R10} & \textbf{R11} & \textbf{R12} \\
\hline
By Color & \checkmark & \checkmark & \checkmark & \checkmark & \checkmark & \checkmark & \checkmark & \checkmark & \checkmark & \checkmark & \checkmark & \checkmark \\
By Form & & & & & & \checkmark & & & & & & \checkmark \\
By Additional Visual Information & & & & \checkmark & \checkmark & & & \checkmark & & & & \\
\bottomrule
\end{tabular}
\caption{Participants' selection of AR distinction designs in Study I. Participants could combine multiple distinction methods to distinguish primary- and secondary-important objects. All 12 participants used color distinction. Five participants additionally combined it with distinguishing by form (R6, R12) or by additional visual information (R4, R5, R8).}
\Description{A table showing each participant's selection of AR distinction designs. All 12 participants used distinguishing by color. Two participants (R6, R12) additionally used distinguishing by form, and three participants (R4, R5, R8) additionally used distinguishing by additional visual information.}
\label{tab:distinction_selection}
\vspace{-5ex}
\end{table*}

\subsubsection{Preferences on AR Distinction Designs}
\label{subsubsec:indoor_preferences}
Compared to augmenting all important objects equally, eleven participants (all except R6) preferred the AR distinction in \system{} because it helped guide their attention (see Section \ref{subsubsec:attention_allocation}). 
Ten participants (e.g., R2, R7) appreciated that the AR distinction emphasized dangerous objects, allowing them to pinpoint these objects more quickly. Eight participants (e.g., R5, R10) also found that the AR distinction helped with object identification because each augmentation represented a smaller set of objects and narrowed down their identification scope. For example, R1 used yellow outline for primary-important objects and blue outline for secondary ones: ``When I saw something in yellow, I knew that it would have to be one of those [four] items, like the glass or the scissors, and when I saw something in blue I knew it was a bowl, a fork, or a spoon.''
In contrast, R6 preferred augmenting all objects equally, as she found that the AR distinction added cognitive load: ``I think I'm doing too much thinking [with distinction] like, oh, some [augmentations] are blue, why are they blue? So I'm kind of going down that way.''

When choosing distinction designs, all 12 participants chose distinguishing \textit{By Color}, and five combined it with \textit{By Form} (R6, R12) or \textit{By Additional Visual Information} (R4, R5, R8). Table \ref{tab:distinction_selection} lists participants' distinction design choices. Participants chose distinction design based on two criteria: intuitive mapping between augmentation and importance, and avoiding visual clutter.

\textbf{Intuitive mapping between augmentation and importance.}
Participants evaluated the AR distinction designs on how intuitively the visual difference mapped to importance levels. Six participants (e.g., R5, R9) considered \textit{By Color} intuitive and easy to interpret. By assigning a brighter color to primary-important objects (e.g., yellow or green) and a less bright color to secondary ones (e.g., blue), nine participants (e.g., R4, R8) reported that the brighter augmentations stood out and naturally drew their attention at first glance without additional thinking. As R9 said: ``When it was different colors, that was very clear that this is a [primary-important] object, whereas [with the same color] I have to kind of stop and read the icon a little bit, so it's not as immediate.''

When evaluating \textit{By Form} distinction, three participants (R6, R7, R12) found the pairing of solid overlay for primary-important objects and outline for secondary ones intuitive, because both augmentations were rendered on the object and differed only in visual coverage. As R12 explained, the solid overlay ``adds a very clear distinction'' without ``the complexity of adding the vertical lines and the icons.'' In contrast, no participant chose to use outlines for one importance level and icon labels for the other. While these two designs were visually different, they did not intuitively convey importance levels (R5, R6, R9). R5 and R9 also found \textit{By Additional Visual Information} unintuitive to indicate importance.


\textbf{Avoiding visual clutter.}
Three participants (R4, R5, R8) combined \textit{By Color} with \textit{By Additional Visual Information} to highlight primary-important objects. They felt that outline alone did not help with object identification, but adding icon labels to objects would create high visual clutter. Selectively adding icons to only primary-important objects balanced visual clutter and clarity (R2, R4, R5, R8). As R4 described, this method ``is a happy medium of helping me identify, but also not like an insane level of distraction.''
R3 and R9 cautioned that combining multiple augmentations on the same object could create visual clutter, as the augmentations intertwined and became difficult to parse. For example, R3 found that the outlines ``get messed up with the icons,'' making it messy and ``hard to concentrate.''

\section{Study II: Using \system{} in Busy Street Navigation}
\label{sec:outdoor_study}
While Study I revealed the impact of \system{} as well as the challenges of augmenting multiple objects in a complex kitchen scene, \colorchange{it contained only static objects, whereas complex scenes can also involve dynamic objects, a key dimension of complexity absent in Study I.}
Therefore, \colorchange{we supplemented it with an outdoor scenario that involves moving objects}. Specifically, we conducted a think-aloud study with 13 PLV who navigated a preplanned outdoor route using \system{} to explore the unique opportunities and challenges in complex, dynamic outdoor scenes. 
Although prior work proposed head-mounted AR systems for the outdoor mobility of visually impaired people \cite{lo2021navigation, min2021augmented}, to our knowledge, none were evaluated in real-world outdoor scenarios.


\subsection{Methods}

\subsubsection{Participants}
We recruited 13 participants (T1--T13) from the same source as Study I (Section \ref{subsubsec:kitchen_participants}). Table \ref{tab:demographics_outdoor} in Appendix~\ref{app:demographic_outdoor} details participants' demographic information and visual conditions.
All except T1, T8, and T11 were legally blind. Five participants (T3, T5, T8, T10, T12) had low visual acuity (i.e., no better than 20/100), eight (T2--T4, T6, T7, T9, T12, T13) had peripheral vision loss, and four (T5, T8, T11, T13) had central vision loss. Three participants (T3, T6, T9) had prior experience with AR, but none used it regularly. Three participants attended both studies (R1/T3; R6/T5; R8/T10).
Participants were compensated \$25 per hour. The study was approved by our university IRB.

\subsubsection{Apparatus}
\begin{figure}[htbp]
    \centering
    \includegraphics[width=0.8\linewidth]{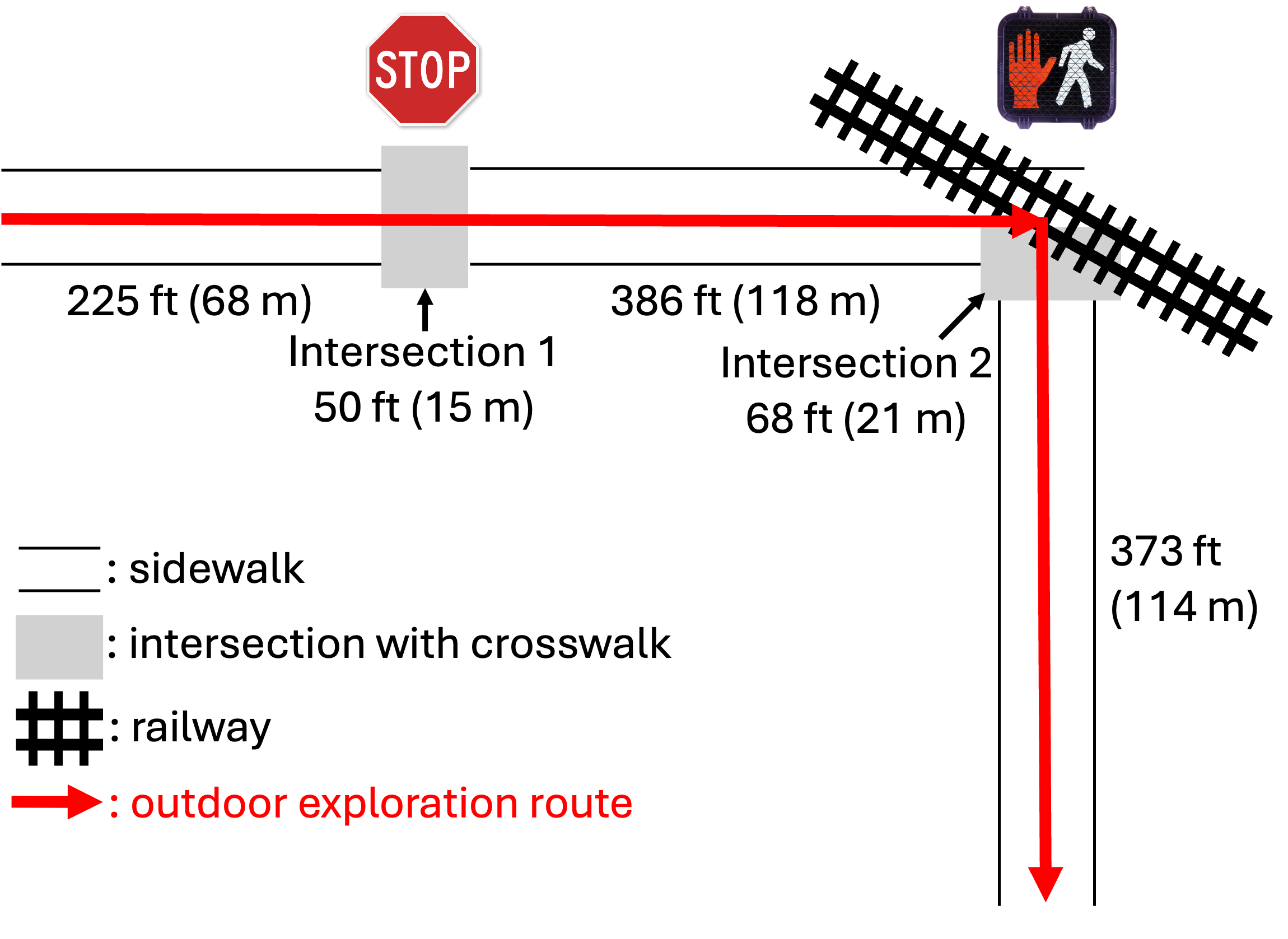}
  \caption{The route for the outdoor navigation study, which crossed two intersections and passed a railway crossing.}
  \Description{The route for the outdoor navigation study. It crossed two intersections and passed a railway crossing.}
  \label{fig:summative_outdoor_route}
\end{figure}

We pre-planned an outdoor route of approximately 1,100 feet (335 meters) long with two intersections. The first intersection was labeled by a stop sign and the second by traffic lights. A railway crossed the route near the second intersection. Participants walked on the sidewalk except when crossing the two intersections. Overall, the route represented a complex walking path for PLV. See Figure \ref{fig:summative_outdoor_route} for an overview of the route.

\subsubsection{Procedure}
Due to the difficulty and potential safety risks of conducting a quantitative outdoor navigation experiment, we conducted a free-form think-aloud study to gain a preliminary understanding of PLV's experiences and challenges with \system{}. 
The study lasted approximately one hour. Participants navigated along the preplanned route using \system{} (Figure \ref{fig:summative_outdoor_route}). At the beginning of the route, participants tried out all base augmentations and AR distinction designs, customized them, and selected their preferred designs. They were allowed to not use AR distinction but only one base augmentation if they preferred due to safety consideration.
Then, participants walked along the route with \system{} using their preferred augmentations. During the walk, participants ``thought aloud,'' discussing their experience with \system{}, including benefits, challenges, and suggestions for improvement. One researcher on the team walked with them to ensure safety \colorchange{and a second researcher video-recorded the session for later analysis}.

We followed the same procedure as in Study I to analyze the qualitative data (Section \ref{subsubsec:kitchen_qualitative_analysis}): \colorchange{two researchers open-coded three participants' transcripts (23\% of the data) to develop an initial codebook upon agreement. One researcher then coded the remaining transcripts and derived themes on participants' experiences with \system{}, including its benefits in outdoor navigation, the challenges of perceiving AR-enhanced dynamic scenes, and participants' suggestions and design preferences.}

\subsection{Findings}
We identified two unique scenarios in complex outdoor environments, where \system{} was helpful but also encountered augmentation challenges: perceiving ground surfaces and dynamic objects. \colorchange{We also found that different recognition errors (false negatives vs. false discoveries) affected participants' safety asymmetrically.} Moreover, participants' preferences on AR distinction designs changed due to outdoor lighting conditions. We elaborate on these findings below.

\subsubsection{Perceiving ground surfaces}
\label{subsubsec:augmenting_ground_surface}
Unlike the kitchen environment where augmented objects were small, discrete, and bounded (e.g., bowls, knives, glasses), the outdoor scene contained large, continuous ground surfaces (e.g., sidewalks, crosswalks, railway tracks). Participants reported that augmenting these surfaces supported safer navigation. Ten participants (e.g., T1, T8) appreciated that the AR outline clarified sidewalk and crosswalk boundaries and helped them stay within walkable areas. Augmenting walkable surfaces also helped participants better notice and avoid tripping hazards, such as curbs (11/13; e.g., T2, T13) and railway crossings (7/13; e.g., T3, T7). As T4 said: ``[The outline] helps you with taking your path [...] [knowing] the sidewalk and where you can't go. That makes you aware that: okay, that's the [sidewalk] edge there, or there's an up or down [curb] that could trip you.''

Since tripping hazards were categorized as primary-important objects, the AR distinction further helped participants attend to them first among all augmented objects (6/13; e.g., T8, T11). As T12 explained: ``I think having the differentiation kind of outlined what was more important [...] you can only take in so much information and you want to take in the most relevant stuff first. Like the fact that there's a curb coming, that's important so you don't fall.''

\begin{figure*}[tbp]
    \centering
    \includegraphics[width=\textwidth]{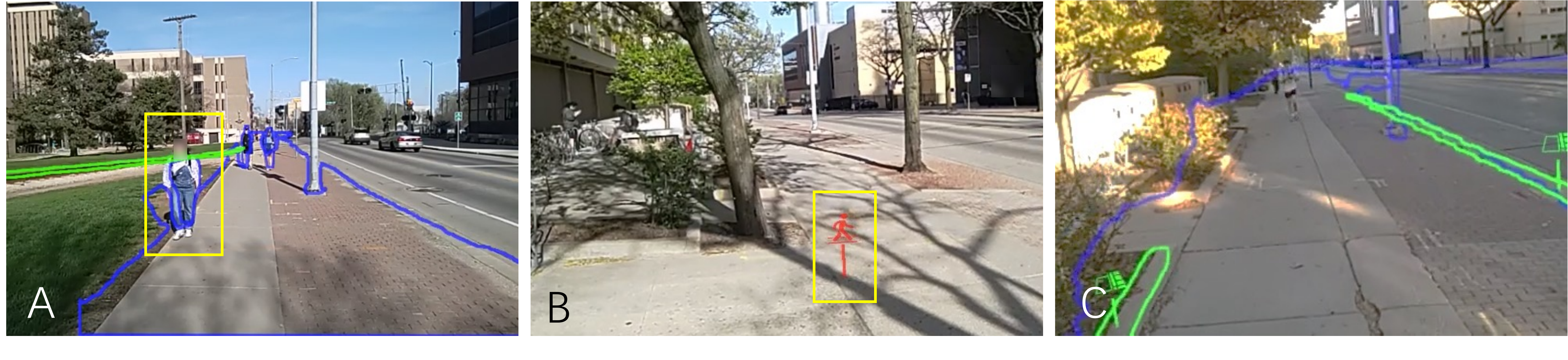}
  \caption{Examples of perception challenges identified in Study II. (A) Outlines of the sidewalk and a railway track crossed the outline of a pedestrian, creating visually confusing shapes. (B) An icon label on a walkable surface did not indicate which area it was augmenting. (C) Outlines of the sidewalk squiggled rather than precisely tracing the sidewalk edge.}
  \Description{This figure contains three sub-figure, corresponding to the three design challenges of augmenting ground surfaces in outdoor environments. (A) Outlines of the sidewalk and a railway track crossed the outline of a pedestrian standing on the sidewalk, creating visually confusing shapes. (B) An icon label on a walkable surface did not indicate which area it was augmenting, as it was pointing to a random spot on the surface. (C) Outlines of the sidewalk squiggled rather than precisely tracing the sidewalk edge.}
  \label{fig:outdoor_problem}
\end{figure*}

Despite its promise, participants reported several augmentation challenges associated with ground surfaces:

\textbf{Augmentations interfered with objects on the surface.}
Because ground surfaces stretched continuously across the scene, the AR outlines frequently intersected with other objects on the surface and their augmentations (e.g., people, poles), creating confusing shapes (5/13; e.g., T3, T13). For example, T13 felt confused when the sidewalk outline crossed the outline of a pedestrian (Figure \ref{fig:outdoor_problem}A). The solid overlay exacerbated this problem, as overlays of surfaces and overlays of objects on them merged into one colored shape (5/13; e.g., T8, T9). While the AR distinction helped reduce this visual interference (T3), it did not work when the surface and objects on it had the same importance level. T11 thus suggested using a designated color for walkable surfaces to visually separate them from other augmented objects.

\textbf{Icon labels did not convey surface boundaries.}
While icon labels avoided the intersection problem, they did not convey surface boundaries, which were critical information for navigation (T7--T9, T11). Moreover, without explicitly highlighting surface boundaries, three participants (T5, T8, T11) found it hard to understand what the icons were referring to (Figure \ref{fig:outdoor_problem}B). As T8 noted: ``I think [the icon] is cool, but I think I like the outlines better. Just being outdoors in the wide open space, it's nicer to have the outlines, whereas [with the icon] it's one object kind of defining a lot more of the area.'' T7 suggested combining icon labels with outlines to both identify surfaces and convey their boundaries.

\textbf{Amplified impact of object segmentation errors on large surfaces.}
Since ground surfaces were large and extended across the scene, the inaccuracies in the mask from the object segmentation model were amplified and appeared obvious. As a result, six participants (e.g., T6, T9) reported that the AR outlines of sidewalks and crosswalks were confusing because they squiggled rather than precisely tracing the surface edge (Figure \ref{fig:outdoor_problem}C).
As T9 said: ``[\system{}] doesn't tell you exactly what the sidewalk is. It just gives me some lines, some squiggly lines, some straight lines [...] so it's not really accurate.'' T12 also said that she ``wouldn't rely on [surface augmentations] for safety'' due to the inaccurate outlines. Since the icon labels were placed at the center of the segmentation masks, T5 also found the icons for the sidewalk constantly ``running around'' as she walked, bringing more confusion.

\subsubsection{Perceiving and distinguishing dynamic objects}
\label{subsubsec:dynamic_objects}
Unlike the static kitchen environment, the outdoor scene contained dynamic objects (e.g., pedestrians and cyclists). Participants reported that augmentation helped them perceive these moving objects, and suggested that \system{} should distinguish dynamic objects based on the likelihood of \colorchange{collision}. 

\textbf{Augmentations revealed object motion.}
Ten participants (e.g., T1, T9) reported that augmentation helped them perceive dynamic objects such as pedestrians and cyclists. Because the augmentations moved together with the dynamic objects, participants could notice when an object was moving and judge its direction. As T9 described: ``[From the augmentation] you can tell that [pedestrians] are walking. They're walking towards me. You can tell by their legs when they're moving.'' Seven participants (e.g., T4, T10) further noted that this awareness helped them anticipate potential collisions and plan ahead to avoid risks. 

\textbf{Importance depended on \colorchange{path-crossing likelihood}.}
While participants appreciated augmenting moving objects, seven participants (e.g., T1, T3) also suggested that the importance level of dynamic objects should depend on \colorchange{their likelihood of crossing the user's path}. Objects more likely to \colorchange{cross their paths}, such as cars at crosswalks (T3, T8, T10) or incoming pedestrians on the same sidewalk (T3, T7, T8), should have higher importance so that participants could notice them quickly. Objects less likely to do so, such as moving cars on the road while walking on the sidewalk (T1, T4, T7), should be of lower importance or not augmented. As T4 described: ``with walking on the sidewalk [...] I don't think I need to see the cars as much [...] But getting to the crosswalk, augmenting the cars might be helpful.''

\subsubsection{Asymmetric Impact of Recognition Errors on Safety}
\colorchange{
Beyond the augmentation challenges, participants' safety also depended on the recognition accuracy. We found that recognition errors affected safety asymmetrically. False negatives were safety-critical, since missing an important object would leave participants unaware of a hazard (T7, T9). For example, when T9 approached the other side of a crossing, neither the curb nor the curb cut was augmented, so she could not tell where to step onto the sidewalk without tripping over the curb. In contrast, false discoveries were less safety-critical and mainly caused confusion (T9, T11), such as seeing an augmentation for a person who was not there (T9). T2 even found false discoveries helpful: when a bike rack (not in the original important object list) was augmented as a fence, the augmentation made her aware of the bike rack so she could avoid it.
}

\begin{table*}[t]
\centering
\small
\begin{tabular}{p{4cm}ccccccccccccc}
\toprule
\textbf{AR Distinction Design} & \textbf{T1} & \textbf{T2} & \textbf{T3} & \textbf{T4} & \textbf{T5} & \textbf{T6} & \textbf{T7} & \textbf{T8} & \textbf{T9} & \textbf{T10} & \textbf{T11} & \textbf{T12} & \textbf{T13} \\
\hline
By Color & & \checkmark & \checkmark & \checkmark & & \checkmark & \checkmark & \checkmark & \checkmark & & \checkmark & & \\
By Form & & & & & \checkmark & & & \checkmark & & & & \checkmark & \\
By Additional Visual Information & & \checkmark & & & & \checkmark & \checkmark & & & & & & \checkmark \\
By Outline Thickness & \checkmark & & & & & & & & & & & & \\
No Distinction & & & & & & & & & & \checkmark & & & \\
\bottomrule
\end{tabular}
\caption{Participants' selection of AR distinction designs in the outdoor study. Unlike the kitchen study where all 12 participants chose distinction by color, outdoor participants' choices were more varied. Eight of 13 used color distinction, three used form, and four used additional visual information. T10 preferred no distinction and augmented all important objects using icon labels. T1 proposed distinguishing by outline thickness during the study (i.e., a thicker outline for primary-important objects and a thinner outline for secondary ones).}
\Description{A table showing each participant's selection of AR distinction designs in the outdoor study. Eight participants used distinction by color. Three participants used distinction by form. Four participants used distinction by additional visual information. One participant (T1) proposed distinguishing by outline thickness during the study. One participant (T10) preferred no distinction.}
\label{tab:distinction_selection_outdoor}
\vspace{-3ex}
\end{table*}

\subsubsection{Preferences on AR distinction designs}
Unlike the indoor scenario in Study I, where all participants chose the \textit{By Color} distinction, participants' preferences outdoors appeared more diverse: \colorchange{only eight of the 13 participants chose \textit{By Color}, with T1 proposing a new design (\textit{By Outline Thickness}) and T10 preferring no distinction at all.} 
 Table \ref{tab:distinction_selection_outdoor} lists participants' selection of distinction designs in the outdoor environment.

\colorchange{While applying similar augmentation selection criteria to Study I (Section \ref{subsubsec:indoor_preferences})---the intuitiveness of mapping between augmentation and importance (5/13; e.g., T1, T7) and the severity of visual clutter (6/13; e.g., T4, T8)---participants further} reported that their preferences were affected by the outdoor lighting conditions. \colorchange{Unlike the uniform lighting of the indoor kitchen, which made color differences easy to perceive, the variable outdoor lighting distorted the augmentation colors.}
Eight participants (e.g., T4, T11) found that 
color differences were more subtle and sometimes hard to perceive. For example, T11 used red icon labels for primary-important objects and white icon labels for secondary ones but found that the white icons appeared ``pinky purple'' and looked similar to the red icons when viewed under the sun. As a result, four participants (T7, T8, T12, T13) preferred the \textit{By Form} or \textit{By Additional Visual Information} AR distinction as these designs distinguished objects by shape and remained visible with diminished color contrast. As T8 described, when augmenting primary-important objects with solid overlay and secondary ones with outline, the solid overlay appeared ``very distinguished compared to just the outline of the sidewalk or the outline of the person,'' whereas distinction by color ``kind of blended in together.'' 
\colorchange{Similarly, T1's \textit{By Outline Thickness} (a thicker outline for primary-important objects and a thinner one for secondary ones)}
remained visually distinct in the outdoor environment.

\section{Discussion}
In this paper, we explored the opportunities and challenges of AR augmentation and distinction in supporting PLV's perception of complex scenes. We designed \system{}, a wearable AR system that recognizes, locates, and augments multiple important objects in 3D space with AR distinction.
Through a well-controlled study at a mock-up kitchen counter with 12 PLV, we found that \system{} shifted participants' attention toward primary-important objects. 
However, we also identified perceptual challenges from multi-object augmentations in complex scenes, including perceiving partially occluded and visually similar objects, adjacent augmentations creating visual confusion, and spatial misalignment of icon labels.
A second free-form think-aloud study with 13 PLV in the outdoor street environment further revealed design challenges and needs specific to dynamic, complex scenes, including augmenting continuous ground surfaces and distinguishing dynamic objects by \colorchange{collision} likelihood.
In this section, we discuss the design implications for multiple object augmentations in complex scenes, as well as limitations and future directions.


\subsection{Design Implications for Multiple Object Augmentations in Complex Scenes}
\label{subsec:design_implications}
Summarizing participants' experiences using \system{} across indoor and outdoor scenes, we discuss design implications for future AR systems that augment multiple objects in complex scenes.

\textbf{Distinguish objects by spatial relations.}
Scene complexity is influenced not only by the number of objects but also by their spatial arrangement \cite{oliva2004identifying, kyle2023characterising}. During the kitchen perception task, augmentations on adjacent objects of the same importance level intersected and merged into confusing shapes (Section \ref{subsubsec:perceptual_challenges}). This pattern is consistent with the Gestalt principle of uniform connectedness \cite{palmer1994rethinking, peterson2013gestalt} that adjacent regions with the same visual properties would form a connected uniform region and tend to be perceived as a single shape rather than as distinct objects. \system{}'s distinction did not prevent this problem because it evaluated object importance based only on per-object properties (e.g., risk severity, visual difficulty), not on inter-object relations.
Future research should explore modeling spatial relations between objects (e.g., adjacency, overlap, and containment) when distinguishing objects. For example, systems could integrate scene graph generation \cite{gu2023conceptgraphsopenvocabulary3dscene, chang2021comprehensive} to model objects' spatial relations, or CV techniques such as amodal instance segmentation \cite{li2016amodal} to predict the occluded parts of objects. Given such spatial relations, systems could break the uniformity of adjacent same-importance augmentations by varying a visual dimension, such as color, brightness, texture, or motion \cite{palmer1994rethinking}. \colorchange{Beyond distinguishing overlapping objects, future systems could register static objects and continue augmenting them when partially or fully occluded (e.g., using dotted outlines) to avoid misleading blended shapes (Section~\ref{subsubsec:perceptual_challenges}) and keep hidden objects visible \cite{elmqvist2008taxonomy, macedo2021occlusion}.}

\textbf{Support anchor-based scene perception.}
While the formative study characterized factors of object importance (risk severity, visual difficulty), these factors did not fully capture the perceptual roles some objects played during PLV's perception of complex scenes. During the kitchen perception task, some participants actively looked for specific objects as anchors and organized other objects relative to them when building mental maps (Section \ref{subsubsec:bare_eye_strategy}). They selected anchors based on visual salience, uniqueness in the scene, or safety concern, which only partially aligned with the importance factors used by \system{}. Some anchor objects, such as bowls and cutting boards, were categorized as secondary- or non-important because they were visually obvious, and therefore received less prominent augmentation or were not augmented at all. However, these objects still played a central role in PLV's perception and may warrant visual support. 
Prior work has investigated perceptual landmarks at the scale of buildings, such as when understanding building layouts \cite{kalia2008learning} and conducting indoor navigation \cite{chen2025visimark}, but it remains underexplored how PLV would select and use anchor objects in dense, complex scenes. Future research could investigate what anchors PLV select in different complex scenarios, and visually distinguish them to support anchor-based scene perception.

\textbf{Predict trajectory of dynamic objects.}
Our outdoor findings suggest that augmenting dynamic objects not only made their motion visible but also gave PLV advance notice of potential conflict. Participants appreciated augmentation on dynamic objects (e.g., pedestrians, cyclists) because it helped them anticipate conflicts and react earlier (Section \ref{subsubsec:dynamic_objects}). Seven participants further suggested that the importance of dynamic objects should depend on their likelihood of \colorchange{crossing the user's path}: a car was important at a crosswalk but not while walking on the sidewalk, and an incoming pedestrian on the same sidewalk was more important than one across the street. Future research should predict which objects will soon be relevant from their trajectories and the user's path, such as by incorporating pedestrian trajectory prediction techniques \cite{alahi2016social, gupta2018social}. Given such predictions, future systems could adapt augmentation to give advance notice, such as augmenting with increasing salience (e.g., brighter color, flashing outline) as an object approaches, or showing the object direction with an arrow \cite{kettle2022augmented}. \colorchange{In safety-critical cases, such as fast-approaching or suddenly appearing objects, future systems could add audio or tactile alerts alongside visual cues to convey urgency \cite{gao2025wearable, hersh2022wearable}.}

\colorchange{\textbf{Treat importance as context-dependent rather than object-intrinsic.}
Our findings suggest that importance is context-dependent rather than a fixed, per-object property. While the formative study derived importance from object-level features (risk severity, visual difficulty), Study II participants evaluated importance by the likelihood of collision (Section~\ref{subsubsec:dynamic_objects}). Importance also depended on scene familiarity: in the formative study, a participant wanted only out-of-place objects augmented in familiar settings since he knew where most objects were. These findings suggest that importance is shaped not only by an object's properties but also by the user's situation and prior experience with a scene, echoing prior work showing that attention in real-world scenes is guided by context and prior knowledge \cite{vo2019reading}. Future systems should model importance not only from object properties but also from the user's task, path, and familiarity with the scene \cite{lindlbauer2019context}, such as treating a car as primary-important at a crosswalk but non-important while the user walks on a sidewalk, or tracking a user's visit history to prioritize unfamiliar and out-of-place objects.}



\textbf{Adapt augmentation granularity.}
Participants relied on grouping strategies to perceive complex scenes. During the kitchen perception task without augmentation, participants grouped nearby objects of the same category and memorized these groups rather than individual objects (Section \ref{subsubsec:bare_eye_strategy}). With \system{}, participants also used the spatial distribution of augmentations to build a rough impression of where important objects were concentrated before perceiving individual objects (Section \ref{subsubsec:attention_allocation}). Consistent with prior work showing that grouping by proximity and similarity can support visual working memory \cite{peterson2013gestalt, palmer1994rethinking}, this pattern suggests that PLV may group similar and nearby visual elements into larger units to manage the visual complexity of dense scenes. 
Adaptive information density techniques in AR have long been studied \cite{tatzgern2016adaptive, lindlbauer2019context, park2025exploring}, and LLM-driven approaches have recently enabled contextually adaptive mixed-reality layouts \cite{li2024situationadapt}, but none have extended them to PLV. Future systems for PLV could adapt these techniques to adjust augmentation granularity based on the user's current focus.
For example, for scene overview, the system could merge nearby objects of the same category into a single aggregate augmentation, such as a group outline with a count label (e.g., ``four glasses''). When users focus on a region, the aggregate could expand into per-object augmentations to augment individual objects. Users' focused region could be inferred from head orientation or gaze direction \cite{kyto2018pinpointing, wang2025characterizing}, or from explicit input such as voice or gesture \cite{zhao2019designing}.

\subsection{Limitations and Future Directions}
Our research has three main limitations. 
First, although we involved 22 participants in the two studies exploring design opportunities and challenges, we had only 12 participants for the well-controlled lab study (Study I), limiting the statistical power of our quantitative results. Future studies should involve more participants across diverse visual conditions to fully understand the effects of multi-object augmentation on scene perception for PLV.
\colorchange{Second, three participants enrolled in both studies, potentially causing carryover effects, affecting their AR augmentation choices in the second study. The order of design presentation in the tutorial could also affect participant choices. While we encouraged all participants to freely explore and adjust every design until satisfied before the formal trials to minimize such effects, we could not fully rule them out. Future research could counterbalance the order of designs in the tutorial, and recruit independent samples across studies to further reduce these effects.}
Third, we conducted a free-form think-aloud study (Study II) instead of a controlled experiment for the outdoor environment due to safety risks and the difficulty of controlling outdoor environments. Future work should consider conducting a well-controlled study in complex outdoor scenes with quantitative measures to rigorously assess the benefits and challenges of multi-object augmentation in dynamic environments.
\section{Conclusion}
In this paper, we explored the opportunities and challenges of augmenting and distinguishing multiple objects in complex scenes for people with low vision. We designed \system{}, a wearable AR system that augments and visually distinguishes objects in 3D space based on their perceived importance, and conducted two studies to surface PLV's perception strategies, preferences, and challenges. A well-controlled lab study with 12 PLV at a mock-up kitchen counter showed that \system{} effectively shifted participants' attention toward more important objects, and also surfaced perception challenges specific to multi-object augmentation, including perceiving partially occluded and visually similar objects, merging of adjacent augmentations, and spatial misalignment of icon labels. A second free-form think-aloud study with 13 PLV in outdoor scenes further revealed challenges unique to complex outdoor scenes, such as augmenting continuous ground surfaces, and highlighted the need to distinguish dynamic objects by \colorchange{their likelihood of collision with the user}. Based on these findings, we derive design implications for future AR vision enhancement systems in the complex real world.


\begin{acks}
Research reported in this publication was supported in part by the National
Eye Institute of the National Institutes of Health under Award
Number~R01EY037100. The content is solely the responsibility of
the authors and does not necessarily represent the official views of
the National Institutes of Health.
This work was also supported in part by an Apple Seed Grant and the McPherson Eye Research Institute Grant Accelerator Program at the University of Wisconsin-Madison.
We thank all participants for their time and insights.
\end{acks}

\bibliographystyle{ACM-Reference-Format}
\bibliography{main}

\newpage
\appendix
\onecolumn
\raggedbottom

\section{Design Probes in the Formative Study}
\label{sec:appendix_formative_probe}

\begin{figure}[H]
    \centering
    \includegraphics[width=\linewidth]{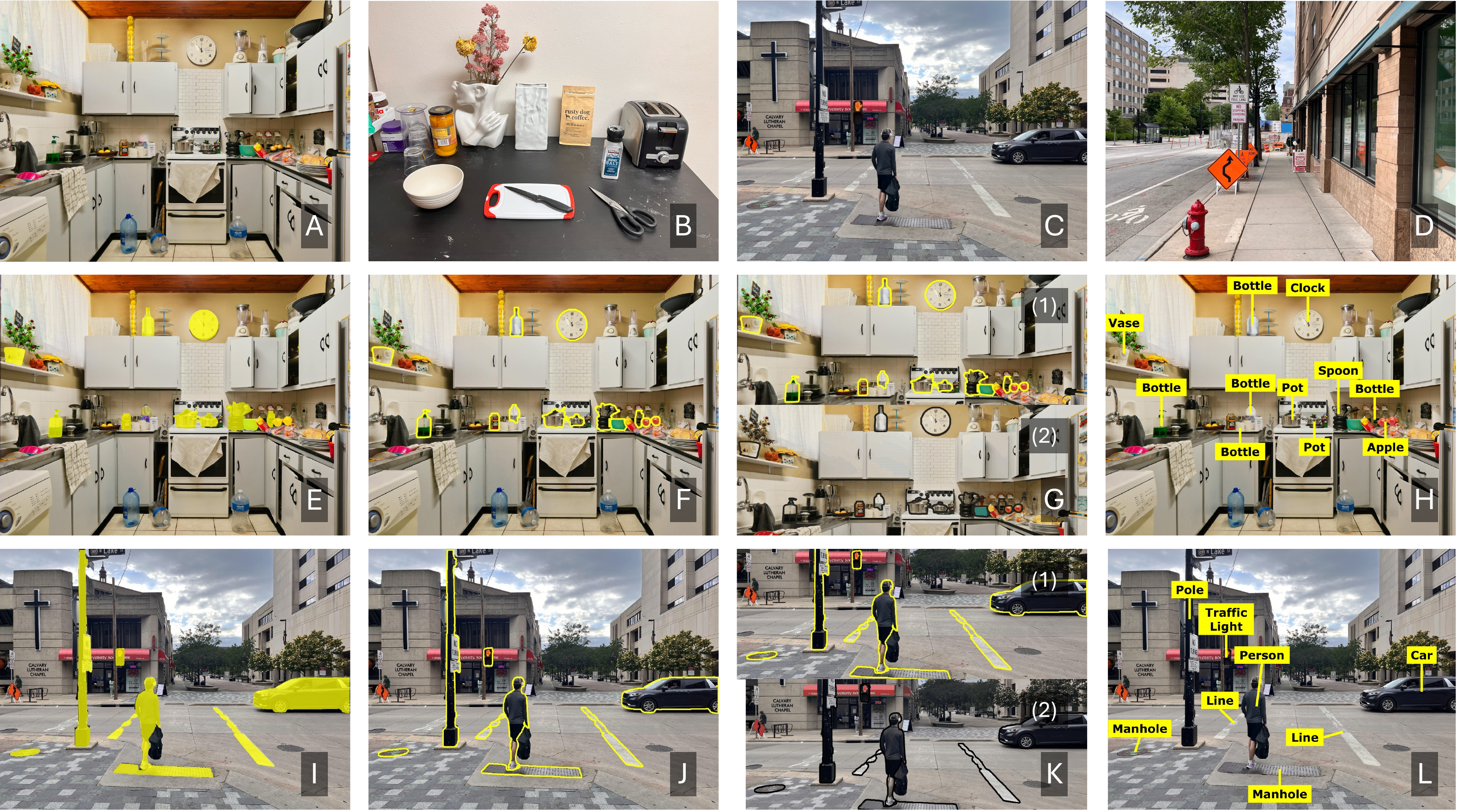}
  \caption{Example scenarios and augmentation designs in the formative study design probe. (A)-(B) Two example kitchen scenarios; (C)-(D) Two example outdoor scenarios; (E) The solid overlay in kitchen with randomly selected objects (vase, bottle, clock, pot, spoon, apple); (F) The static outline in kitchen; (G1-2) The flashing outline in kitchen; (H) The text label in kitchen; (I) The solid overlay in outdoor environment with randomly selected objects (pole, traffic light, person, crosswalk line, manhole, car); (J) The static outline in outdoor environment; (K1-2) The flashing outline in outdoor environment; (L) The text label in outdoor environment.}
  \Description{
   4-by-3 grid of 12 images showing example scenes and AR augmentation designs used as design probes in the formative study. The top row shows four unaugmented example scenes: (A) a kitchen with cabinets, stovetop, and countertop items; (B) a kitchen counter with bowls, utensils, and food items; (C) an outdoor sidewalk scene with a pole and pedestrian; (D) an outdoor street scene with a crosswalk, orange construction cones, and a car.
The middle row shows kitchen augmentation designs applied to randomly selected objects (vase, bottle, clock, pot, spoon, apple): (E) Solid Overlay, with semi-transparent yellow overlays covering each object; (F) Static Outline, with bright yellow contours around each object; (G1–2) Flashing Outline, shown in two frames that alternate between visible and invisible outlines to illustrate the flashing effect; (H) Text Label, with yellow rectangular labels containing each object's name floating above the object.
The bottom row shows the same four augmentation designs applied to a busy outdoor scene with randomly selected objects (pole, traffic light, person, crosswalk line, manhole, car): (I) Solid Overlay with semi-transparent yellow overlays on each object; (J) Static Outline with yellow contours around each object; (K1–2) Flashing Outline shown as two alternating frames; (L) Text Label with yellow rectangular name labels floating above each object.
  }
  \label{fig:design_probe_1}
\end{figure}

\begin{figure}[H]
    \centering
    \includegraphics[width=\linewidth]{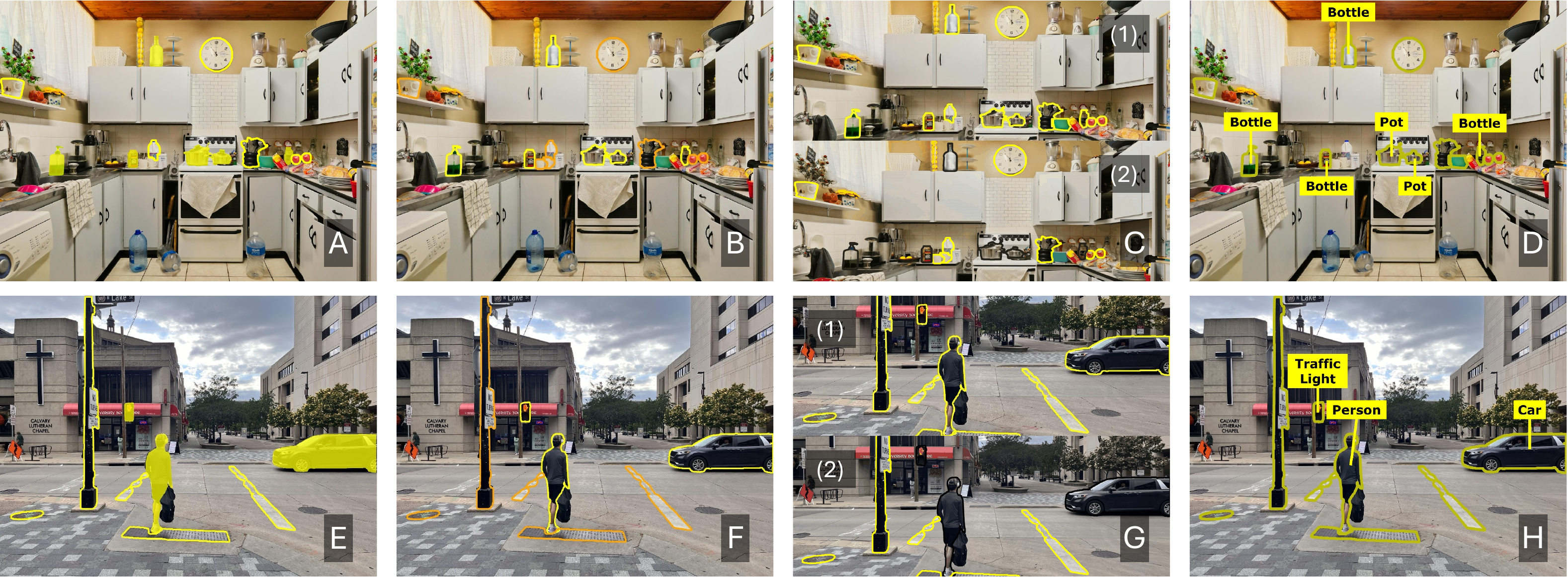}
  \caption{Example distinction methods in the formative study design probe. (A) The \textit{By Form} in kitchen with randomly selected high-importance objects (bottle, pot) in solid overlay and low-importance objects (vase, spoon, clock, apple) in static outline; (B) The \textit{By Color} in kitchen with high-importance objects in yellow static outline and low-importance ones in orange static outline; (C1-2) The \textit{By Visual Effect} in kitchen with high-importance objects in flashing outline and low-importance ones in static outline; (D) The \textit{By Additional Visual Information} in kitchen with high-importance objects in static outline and text label combined, and low-importance ones in static outline only; (E) The \textit{By Form} in outdoor environment with randomly selected high-importance objects (person, car, traffic light) in solid overlay and low-importance ones (manhole, crosswalk line, pole) in static outline; (F) The \textit{By Color} in outdoor environment; (G1-2) The \textit{By Visual Effect} in outdoor environment; (H) The \textit{By Additional Visual Information} in outdoor environment.}
  \Description{
  A 2-by-4 grid of images showing the four AR distinction methods used as design probes in the formative study, each illustrated in both a kitchen and an outdoor setting. The top row shows the four methods applied in a kitchen scene, with primary-importance objects (bottle, pot) and secondary-importance objects (vase, spoon, clock, apple) augmented differently: (A) By Form, with primary objects rendered in yellow solid overlay and secondary objects in yellow static outline; (B) By Color, with primary objects in bright yellow static outline and secondary objects in darker orange static outline; (C1–2) By Visual Effect, shown as two alternating frames in which primary objects are rendered with a flashing outline while secondary objects remain in static outline; (D) By Additional Visual Information, with primary objects rendered in static outline combined with a floating yellow text label, and secondary objects in static outline only.
The bottom row shows the same four methods applied in an outdoor scene, with primary-importance objects (person, car, traffic light) and secondary-importance objects (manhole, crosswalk line, pole) differentiated using the corresponding visual treatments: (E) By Form; (F) By Color; (G1–2) By Visual Effect, shown as two alternating frames; (H) By Additional Visual Information.
  }
  \label{fig:design_probe_2}
\end{figure}

\section{Demographic Information of Formative Study Participants}
\label{app:demographic_formative}

\begin{table}[htbp]
\footnotesize
\centering
\setlength{\tabcolsep}{3pt}
\begin{tabular}{>{\centering\arraybackslash}p{0.55cm}>{\centering\arraybackslash}p{0.65cm}>{\centering\arraybackslash}p{2.9cm}>{\centering\arraybackslash}p{0.85cm}>{\centering\arraybackslash}p{2.4cm}>{\centering\arraybackslash}p{3.2cm}>{\centering\arraybackslash}p{4.0cm}>{\centering\arraybackslash}p{1.15cm}} 
\toprule
  \textbf{ID} & \textbf{Age/} \newline \textbf{Gender} &  \textbf{Diagnosis} & \textbf{Legally Blind} &  \textbf{Visual Acuity} &  \textbf{Field of View} &  \textbf{Other Visual Difficulties} & \textbf{Prior AR \newline Experience} \\
\hline
\multirow{2}{*}{P1} & \multirow{2}{*}{55/M} & \multirow{2}{*}{Stargardt disease} & \multirow{2}{*}{N} & \multirow{2}{*}{L: 20/100; R: 20/75} & Blurry and hazy spots & Sensitive to light & \multirow{2}{*}{N} \\
& & & & &  on central vision & Cannot distinguish between dark colors & \\
\hline
\multirow{2}{*}{P2} & \multirow{2}{*}{58/M} & \multirow{2}{*}{Optic neuropathy} & \multirow{2}{*}{Y} & \multirow{2}{*}{L: 20/70; R: 20/300} & \multirow{2}{*}{Both eyes <19\degree} & \multirow{2}{*}{Sensitive to light} & \multirow{2}{*}{Y} \\
& & & & & & & \\

\hline
\multirow{2}{*}{P3} & \multirow{2}{*}{51/F} & Retinopathy of prematurity, & \multirow{2}{*}{Y} & \multirow{2}{*}{L: 20/300; R: 20/125} & Peripheral vision loss & \multirow{2}{*}{Sensitive to light} & \multirow{2}{*}{N} \\
& & Chronic uveitis & & & Full central vision & & \\

\hline
\multirow{2}{*}{P4} & \multirow{2}{*}{85/F} & Macular Edema, & \multirow{2}{*}{N} & Unknown, but with & Peripheral vision loss & \multirow{2}{*}{Sensitive to light} & \multirow{2}{*}{N} \\
& & Glaucoma & & blurry central vision & Full central vision & & \\

\hline
\multirow{2}{*}{P5} & \multirow{2}{*}{62/F} & \multirow{2}{*}{Spinal meningitis} & \multirow{2}{*}{Y} & \multirow{2}{*}{L: 20/2200; R: 20/400} & Cannot see lower half & \multirow{2}{*}{Sensitive to light} & \multirow{2}{*}{Y} \\
& & & & & of vision & & \\

\hline
\multirow{2}{*}{P6} & \multirow{2}{*}{85/F} & Macular degeneration, & \multirow{2}{*}{Unknown} & Unknown, but with & \multirow{2}{*}{Blurry central vision} & \multirow{2}{*}{Occasional optical illusions} & \multirow{2}{*}{N} \\
& & Charles Bonnet syndrome & & blurry central vision & & & \\
\bottomrule
\end{tabular}
\caption{Participant demographics and visual conditions in the formative study.}
\Description{This table lists out the demographic information and visual conditions of the six low vision participants in the formative study.}
\label{tab:demographics_formative}
\vspace{-5ex}
\end{table}

\section{Per-Category Recognition Accuracy of the Fine-Tuned Models}
\label{app:recognition_accuracy}

\begin{table}[H]
  \centering
  \setlength{\tabcolsep}{4pt}
  \begin{tabular}{@{}l r rr rrr | rrr@{}}
    \toprule
    \multirow{3}{*}{\textbf{Class}} & \multirow{3}{*}{\textbf{Instances}} & \multicolumn{5}{c|}{\textbf{\textit{RTMDet-Ins-l-Kitchen} (fine-tuned)}} & \multicolumn{3}{c}{\textbf{\textit{RTMDet-Ins-l} (baseline)}} \\
    \cmidrule(lr){3-7} \cmidrule(lr){8-10}
     & & \makecell[r]{\textbf{False Negative}\\\textbf{Rate}} & \makecell[r]{\textbf{False Discovery}\\\textbf{Rate}} & \textbf{AP@50} & \textbf{AP@75} & \textbf{AP} & \textbf{AP@50} & \textbf{AP@75} & \textbf{AP} \\
    \midrule
    Bowl      &   626 & 0.364 & 0.365 & \textbf{0.628} & \textbf{0.492} & \textbf{0.454} & 0.357 & 0.294 & 0.267 \\
    Carafe    &   283 & 0.078 & 0.199 & \textbf{0.887} & \textbf{0.843} & \textbf{0.755} & 0.000 & 0.000 & 0.000 \\
    Cup       &  1926 & 0.268 & 0.270 & \textbf{0.746} & \textbf{0.608} & \textbf{0.538} & 0.489 & 0.388 & 0.344 \\
    Fork      &   616 & 0.263 & 0.210 & \textbf{0.703} & \textbf{0.224} & \textbf{0.316} & 0.441 & 0.128 & 0.193 \\
    Glasses   &   343 & 0.394 & 0.197 & \textbf{0.640} & \textbf{0.423} & \textbf{0.385} & 0.612 & 0.383 & 0.375 \\
    Jars      &  1025 & 0.377 & 0.330 & \textbf{0.644} & \textbf{0.441} & \textbf{0.404} & 0.327 & 0.242 & 0.218 \\
    Knife     &   982 & 0.369 & 0.322 & \textbf{0.582} & \textbf{0.312} & \textbf{0.312} & 0.392 & 0.226 & 0.214 \\
    Ladle     &    61 & 0.344 & 0.394 & \textbf{0.627} & \textbf{0.429} & \textbf{0.373} & 0.000 & 0.000 & 0.000 \\
    Scissors  &   109 & 0.248 & 0.349 & \textbf{0.740} & \textbf{0.599} & \textbf{0.465} & 0.519 & 0.405 & 0.330 \\
    Spatula   &   278 & 0.165 & 0.341 & \textbf{0.750} & \textbf{0.592} & \textbf{0.497} & 0.000 & 0.000 & 0.000 \\
    Spoon     &   929 & 0.351 & 0.294 & \textbf{0.566} & \textbf{0.282} & \textbf{0.288} & 0.304 & 0.135 & 0.149 \\
    \midrule
    \textbf{All classes} & \textbf{7178} & \textbf{0.293} & \textbf{0.297} & \textbf{0.683} & \textbf{0.477} & \textbf{0.435} & 0.313 & 0.200 & 0.190 \\
    \bottomrule
  \end{tabular}
  \caption{Per-category recognition performance of the kitchen model (\textit{RTMDet-Ins-l-Kitchen}) against the baseline (\textit{RTMDet-Ins-l}) on the kitchen test set. \textit{Instances} is the number of ground-truth instances of each class. The \textit{false negative rate} (false negatives / ground truth) indicates the rate of unaugmented important objects, and the \textit{false discovery rate} (false positives / detections) indicates the rate of wrongly augmented non-important objects. For each AP metric, the better model is shown in bold; the fine-tuned model outperforms the baseline on all classes. A baseline score of 0.000 indicates a class absent from MS-COCO (e.g., carafe, ladle, spatula). Detection-level metrics use an IoU threshold of 0.3 and a confidence threshold of 0.45, matching the deployment parameters.}
  \Description{A table with ten columns and twelve rows. The columns are the object class, the number of ground-truth instances, five metrics for the fine-tuned RTMDet-Ins-l-Kitchen model (false negative rate, false discovery rate, AP50, AP75, andAP), and three metrics for the RTMDet-Ins-l baseline (AP50, AP75, and AP). The rows list eleven kitchen object classes in alphabetical order from bowl to spoon, followed by a summary row for all classes. Across all 7,178 instances, the fine-tuned model has a false negative rate of 0.293 and a false discovery rate of 0.297, with AP50 of 0.683, AP75 of 0.477, and AP of 0.435, compared with 0.313, 0.200, and 0.190 for the baseline. The fine-tuned model scores higher than the baseline on every class and every AP metric. Carafe, ladle, and spatula have baseline scores of zero.}
  \label{tab:kitchen_recognition}
\end{table}

\begin{table}[H]
  \centering
  \setlength{\tabcolsep}{4pt}
  \begin{tabular}{@{}l r rr rrr | rrr@{}}
    \toprule
    \multirow{3}{*}{\textbf{Class}} & \multirow{3}{*}{\textbf{Instances}} & \multicolumn{5}{c|}{\textbf{\textit{RTMDet-Ins-l-Street} (fine-tuned)}} & \multicolumn{3}{c}{\textbf{\textit{RTMDet-Ins-l} (baseline)}} \\
    \cmidrule(lr){3-7} \cmidrule(lr){8-10}
     & & \makecell[r]{\textbf{False Negative}\\\textbf{Rate}} & \makecell[r]{\textbf{False Discovery}\\\textbf{Rate}} & \textbf{AP@50} & \textbf{AP@75} & \textbf{AP} & \textbf{AP@50} & \textbf{AP@75} & \textbf{AP} \\
    \midrule
    Bench                      &    47 & 0.511 & 0.395 & \textbf{0.403} & 0.084 & \textbf{0.182} & 0.336 & \textbf{0.119} & 0.175 \\
    Bicycle                    &   195 & 0.364 & 0.205 & \textbf{0.659} & 0.126 & \textbf{0.268} & 0.541 & \textbf{0.127} & 0.222 \\
    Construction cone          &    41 & 0.171 & 0.393 & \textbf{0.781} & \textbf{0.741} & \textbf{0.517} & 0.000 & 0.000 & 0.000 \\
    Crosswalk                  &   277 & 0.791 & 0.293 & \textbf{0.309} & \textbf{0.174} & \textbf{0.173} & 0.000 & 0.000 & 0.000 \\
    Curb                       &  2766 & 0.387 & 0.260 & \textbf{0.517} & \textbf{0.187} & \textbf{0.231} & 0.000 & 0.000 & 0.000 \\
    Curb cut                   &   337 & 0.890 & 0.327 & \textbf{0.150} & \textbf{0.022} & \textbf{0.052} & 0.000 & 0.000 & 0.000 \\
    Fence                      &  1857 & 0.731 & 0.223 & \textbf{0.363} & \textbf{0.173} & \textbf{0.186} & 0.000 & 0.000 & 0.000 \\
    Fire hydrant               &    16 & 0.125 & 0.263 & \textbf{0.864} & \textbf{0.864} & \textbf{0.618} & 0.792 & 0.754 & 0.603 \\
    Lamp pole                  &  4579 & 0.340 & 0.247 & \textbf{0.415} & \textbf{0.038} & \textbf{0.122} & 0.000 & 0.000 & 0.000 \\
    Mailbox                    &    12 & 0.750 & 0.625 & \textbf{0.180} & \textbf{0.077} & \textbf{0.101} & 0.000 & 0.000 & 0.000 \\
    Motorcycle                 &   183 & 0.262 & 0.224 & \textbf{0.752} & \textbf{0.270} & \textbf{0.357} & 0.623 & 0.167 & 0.264 \\
    Pedestrian                 &  1315 & 0.157 & 0.231 & \textbf{0.835} & 0.378 & \textbf{0.435} & 0.738 & \textbf{0.397} & 0.409 \\
    Pedestrian signal$^{*}$    &   124 & 0.387 & 0.420 & \textbf{0.617} & \textbf{0.385} & \textbf{0.357} & 0.030 & 0.017 & 0.016 \\
    Railway track              &    89 & 0.596 & 0.294 & \textbf{0.356} & \textbf{0.165} & \textbf{0.190} & 0.000 & 0.000 & 0.000 \\
    Sewer drain                &    99 & 0.566 & 0.511 & \textbf{0.414} & \textbf{0.195} & \textbf{0.230} & 0.000 & 0.000 & 0.000 \\
    Sidewalk                   &  2615 & 0.391 & 0.202 & \textbf{0.580} & \textbf{0.254} & \textbf{0.286} & 0.000 & 0.000 & 0.000 \\
    Traffic sign               &   908 & 0.227 & 0.371 & \textbf{0.762} & \textbf{0.724} & \textbf{0.588} & 0.106 & 0.099 & 0.086 \\
    Trash can                  &   202 & 0.441 & 0.358 & \textbf{0.590} & \textbf{0.449} & \textbf{0.396} & 0.000 & 0.000 & 0.000 \\
    Utility box                &   177 & 0.582 & 0.426 & \textbf{0.417} & \textbf{0.311} & \textbf{0.259} & 0.000 & 0.000 & 0.000 \\
    Vehicle                    &  6693 & 0.104 & 0.177 & \textbf{0.906} & \textbf{0.790} & \textbf{0.686} & 0.791 & 0.578 & 0.548 \\
    Vehicle signal$^{*}$       &   552 & 0.112 & 0.325 & \textbf{0.845} & \textbf{0.670} & \textbf{0.566} & 0.674 & 0.536 & 0.472 \\
    \midrule
    \textbf{All classes}       & \textbf{23084} & \textbf{0.423} & \textbf{0.322} & \textbf{0.558} & \textbf{0.337} & \textbf{0.324} & 0.221 & 0.133 & 0.133 \\
    \bottomrule
  \end{tabular}
  \caption{Per-category recognition performance of the outdoor model (\textit{RTMDet-Ins-l-Street}) against the baseline (\textit{RTMDet-Ins-l}) on the outdoor test set. \textit{Instances} is the number of ground-truth instances of each class. The \textit{false negative rate} (false negatives / ground truth) indicates the rate of unaugmented important objects, and the \textit{false discovery rate} (false positives / detections) indicates the rate of wrongly augmented non-important objects. For each AP metric, the better model is shown in bold; the fine-tuned model outperforms the baseline on all classes except bench, bicycle, and pedestrian on AP75. A baseline score of 0.000 indicates a class absent from MS-COCO (e.g., sidewalk, curb, curb cut). Detection-level metrics use an IoU threshold of 0.3 and a confidence threshold of 0.45, matching the deployment parameters. $^{*}$MS-COCO has a single \textit{traffic light} class and does not distinguish between pedestrian light and vehicle traffic light, so the same set of baseline detections is evaluated against both classes.}
  \Description{A table with ten columns and twenty-two rows. The columns are the object class, the number of ground-truth instances, five metrics for the fine-tuned RTMDet-Ins-l-Street model (false negative rate, false discovery rate, AP50, AP75, and AP), and three metrics for the RTMDet-Ins-l baseline (AP50, AP75, and AP). The rows list twenty-one outdoor object classes in alphabetical order from bench to vehicle signal, followed by a summary row for all classes. Across all 23,084 instances, the fine-tuned model has a false negative rate of 0.423 and a false discovery rate of 0.322, with AP50 of 0.558, AP75 of 0.337, and AP of 0.324, compared with 0.221, 0.133, and 0.133 for the baseline. The fine-tuned model scores higher than the baseline on every class and metric except bench, bicycle, and pedestrian on AP75. Twelve classes that are absent from MS-COCO have baseline scores of zero.}
  \label{tab:outdoor_accuracy}
\end{table}

\section{Demographic Information of Study I Participants}
\label{app:demographics_kitchen}

\begin{table}[H]
\footnotesize
\centering
\setlength{\tabcolsep}{3pt}
\begin{tabular}{>{\centering\arraybackslash}p{0.55cm}>{\centering\arraybackslash}p{0.65cm}>{\centering\arraybackslash}p{2.9cm}>{\centering\arraybackslash}p{0.85cm}>{\centering\arraybackslash}p{2.4cm}>{\centering\arraybackslash}p{3.2cm}>{\centering\arraybackslash}p{4.0cm}>{\centering\arraybackslash}p{1.15cm}} 
\toprule
  \textbf{ID} & \textbf{Age/} \newline \textbf{Gender} &  \textbf{Diagnosis} & \textbf{Legally Blind} &  \textbf{Visual Acuity} &  \textbf{Field of View} &  \textbf{Other Visual Difficulties} & \textbf{Prior AR \newline Experience} \\

\hline
\multirow{2}{*}{R1} & \multirow{2}{*}{63/F} & \multirow{2}{*}{Spinal meningitis} & \multirow{2}{*}{Y} & \multirow{2}{*}{L: 20/2200; R: 20/400} & Cannot see lower half & \multirow{2}{*}{Sensitive to light} & \multirow{2}{*}{Y} \\
& & & & & of vision & & \\

\hline
\multirow{2}{*}{R2} & \multirow{2}{*}{20/M} & \multirow{2}{*}{Stargardt disease} & \multirow{2}{*}{N} & \multirow{2}{*}{L: 20/200; R: 20/160} & Blurry spots in central vision & \multirow{2}{*}{N/A} & \multirow{2}{*}{N} \\
& & & & & Full peripheral vision & & \\

\hline
\multirow{2}{*}{R3} & \multirow{2}{*}{69/F} & \multirow{2}{*}{Macular dystrophy} & \multirow{2}{*}{Y} & \multirow{2}{*}{L: 20/125; R: 20/160} & Right peripheral vision loss & \multirow{2}{*}{Sensitive to light} & \multirow{2}{*}{N} \\
& & & & & due to bleeding & & \\

\hline
\multirow{2}{*}{R4} & \multirow{2}{*}{29/M} & \multirow{2}{*}{Stargardt disease} & \multirow{2}{*}{N} & \multirow{2}{*}{L: 20/100; R: 20/100} & Blurry spots in central vision & \multirow{2}{*}{N/A} & \multirow{2}{*}{Y} \\
& & & & & Full peripheral vision & & \\

\hline
\multirow{2}{*}{R5} & \multirow{2}{*}{73/M} & \multirow{2}{*}{Macular degeneration} & \multirow{2}{*}{Y} & \multirow{2}{*}{L: 20/160; R: 20/200} & \multirow{2}{*}{Full} & \multirow{2}{*}{N/A} & \multirow{2}{*}{N} \\
& & & & & & & \\

\hline
\multirow{2}{*}{R6} & \multirow{2}{*}{73/F} & \multirow{2}{*}{Cone dystrophy} & \multirow{2}{*}{Y} & \multirow{2}{*}{L: 20/100; R: 20/125} & \multirow{2}{*}{Full} & Cannot distinguish between yellow & \multirow{2}{*}{N} \\
& & & & & & and green; sensitive to light & \\

\hline
\multirow{2}{*}{R7} & \multirow{2}{*}{19/F} & \multirow{2}{*}{Stargardt disease} & \multirow{2}{*}{Y} & \multirow{2}{*}{L: 20/200; R: 20/200} & Reduced central acuity & \multirow{2}{*}{Sensitive to light} & \multirow{2}{*}{N} \\
& & & & & Full peripheral vision & & \\

\hline
\multirow{2}{*}{R8} & \multirow{2}{*}{54/F} & \multirow{2}{*}{Achromatopsia} & \multirow{2}{*}{Y} & \multirow{2}{*}{L: 20/200; R: 20/400} & \multirow{2}{*}{Full} & \multirow{2}{*}{Full color blind} & \multirow{2}{*}{N} \\
& & & & & & & \\

\hline
\multirow{2}{*}{R9} & \multirow{2}{*}{37/M} & \multirow{2}{*}{Partial achromatopsia} & \multirow{2}{*}{Y} & \multirow{2}{*}{L: 20/400; R: 20/400} & \multirow{2}{*}{Full} & \multirow{2}{*}{Difficulty differentiating colors} & \multirow{2}{*}{Y} \\
& & & & & & & \\

\hline
\multirow{2}{*}{R10} & \multirow{2}{*}{57/F} & \multirow{2}{*}{Macular degeneration} & \multirow{2}{*}{N} & \multirow{2}{*}{L: 20/125; R: 20/320} & Dark spots in & \multirow{2}{*}{Sensitive to light} & \multirow{2}{*}{Y} \\
& & & & & central vision & & \\

\hline
\multirow{2}{*}{R11} & \multirow{2}{*}{41/M} & Central scotoma due to & \multirow{2}{*}{Y} & \multirow{2}{*}{L: 20/300; R: 20/400} & Central scotoma & \multirow{2}{*}{Sensitive to light} & \multirow{2}{*}{Y} \\
& & damage in optic nerve & & & Full Peripheral vision & & \\

\hline
\multirow{2}{*}{R12} & \multirow{2}{*}{50/M} & Macular degeneration & \multirow{2}{*}{Y} & L: 20/400 & Hazy spots in central vision & \multirow{2}{*}{Sensitive to light} & \multirow{2}{*}{Y} \\
& & (Best disease) & & R: 20/600---20/800 & Full Peripheral vision & & \\

\bottomrule
\end{tabular}
\caption{Participant demographics and visual conditions in the kitchen observation study \colorchange{(Study I)}.}
\Description{This table lists out the demographic information and visual conditions of the 12 participants in the kitchen observation study (Study 1).}
\label{tab:demographics_kitchen}
\end{table}

\section{Demographic Information of Study II Participants}
\label{app:demographic_outdoor}

\begin{table}[H]
\footnotesize
\centering
\setlength{\tabcolsep}{3pt}
\begin{tabular}{>{\centering\arraybackslash}p{0.55cm}>{\centering\arraybackslash}p{0.65cm}>{\centering\arraybackslash}p{2.9cm}>{\centering\arraybackslash}p{0.85cm}>{\centering\arraybackslash}p{2.4cm}>{\centering\arraybackslash}p{3.2cm}>{\centering\arraybackslash}p{4.0cm}>{\centering\arraybackslash}p{1.15cm}} 
\toprule
  \textbf{ID} & \textbf{Age/} \newline \textbf{Gender} &  \textbf{Diagnosis} & \textbf{Legally Blind} &  \textbf{Visual Acuity} &  \textbf{Field of View} &  \textbf{Other Visual Difficulties} & \textbf{Prior AR \newline Experience} \\

\hline
\multirow{2}{*}{T1} & \multirow{2}{*}{73/M} & \multirow{2}{*}{Genetic optic atrophy} & \multirow{2}{*}{N} & \multirow{2}{*}{L: 20/100; R: 20/80} & \multirow{2}{*}{Full} & \multirow{2}{*}{Cannot distinguish between dark colors} & \multirow{2}{*}{N} \\
& & & & & & & \\

\hline
\multirow{2}{*}{T2} & \multirow{2}{*}{57/F} & \multirow{2}{*}{Retinitis pigmentosa} & \multirow{2}{*}{Y} & \multirow{2}{*}{L: 20/70; R: 20/70} & \multirow{2}{*}{Both eyes <20\degree} & Cannot distinguish between & \multirow{2}{*}{N} \\
& & & & & & blue and green & \\

\hline
\multirow{2}{*}{T3} & \multirow{2}{*}{62/F} & \multirow{2}{*}{Spinal meningitis} & \multirow{2}{*}{Y} & \multirow{2}{*}{L: 20/2200; R: 20/400} & Cannot see lower half & \multirow{2}{*}{Sensitive to light} & \multirow{2}{*}{Y} \\
& & & & & of vision & & \\

\hline
\multirow{2}{*}{T4} & \multirow{2}{*}{64/M} & \multirow{2}{*}{Retinitis pigmentosa} & \multirow{2}{*}{Y} & \multirow{2}{*}{L: 20/40; R: 20/40} & Peripheral vision loss & \multirow{2}{*}{Color blind} & \multirow{2}{*}{N} \\
& & & & & Full central vision & & \\

\hline
\multirow{2}{*}{T5} & \multirow{2}{*}{72/F} & \multirow{2}{*}{Cone dystrophy} & \multirow{2}{*}{Y} & \multirow{2}{*}{L: 20/160; R: 20/160} & Central vision loss & \multirow{2}{*}{Cannot distinguish between dark colors} & \multirow{2}{*}{N} \\
& & & & & Full peripheral vision & & \\

\hline
\multirow{2}{*}{T6} & \multirow{2}{*}{58/M} & \multirow{2}{*}{Optic neuropathy} & \multirow{2}{*}{Y} & \multirow{2}{*}{L: 20/70; R: 20/300} & \multirow{2}{*}{Both eyes <19\degree} & \multirow{2}{*}{Sensitive to light} & \multirow{2}{*}{Y} \\
& & & & & & & \\

\hline
\multirow{2}{*}{T7} & \multirow{2}{*}{41/M} & \multirow{2}{*}{Retinitis pigmentosa} & \multirow{2}{*}{Y} & \multirow{2}{*}{L: 20/40; R: 20/40} & \multirow{2}{*}{Both eyes <15\degree} & \multirow{2}{*}{Sensitive to light} & \multirow{2}{*}{N} \\
& & & & & & & \\

\hline
\multirow{2}{*}{T8} & \multirow{2}{*}{28/M} & \multirow{2}{*}{Stargardt disease} & \multirow{2}{*}{N} & \multirow{2}{*}{L: 20/100; R: 20/100} & Blurry spot in central vision & \multirow{2}{*}{N/A} & \multirow{2}{*}{N} \\
& & & & & Full peripheral vision & & \\

\hline
\multirow{2}{*}{T9} & \multirow{2}{*}{58/F} & \multirow{2}{*}{Retinitis pigmentosa} & \multirow{2}{*}{Y} & \multirow{2}{*}{L: 20/50; R: 20/50} & Peripheral vision loss & Cannot distinguish between & \multirow{2}{*}{Y} \\
& & & & & Full central vision & red, pink, and orange & \\

\hline
\multirow{2}{*}{T10} & \multirow{2}{*}{53/F} & \multirow{2}{*}{Achromatopsia} & \multirow{2}{*}{Y} & \multirow{2}{*}{L: 20/200; R: 20/400} & \multirow{2}{*}{Full} & \multirow{2}{*}{Full color blind} & \multirow{2}{*}{N} \\
& & & & & & & \\

\hline
\multirow{2}{*}{T11} & \multirow{2}{*}{82/F} & \multirow{2}{*}{Macular degeneration} & \multirow{2}{*}{N} & \multirow{2}{*}{L: 20/50; R: 20/60} & L: Central vision loss & \multirow{2}{*}{Sensitive to light} & \multirow{2}{*}{N} \\
& & & & & R: Full & & \\

\hline
\multirow{2}{*}{T12} & \multirow{2}{*}{40/F} & Glaucoma; & \multirow{2}{*}{Y} & L: light perception & R: Peripheral vision loss, & \multirow{2}{*}{Sensitive to light} & \multirow{2}{*}{N} \\
& & injury on right eye & & R: blurry vision & full central vision& & \\

\hline
\multirow{2}{*}{T13} & \multirow{2}{*}{60/M} & Acute asymmetric & \multirow{2}{*}{Y} & \multirow{2}{*}{L: <20/90; R: 20/90} & Scotoma spread out & \multirow{2}{*}{Sensitive to light} & \multirow{2}{*}{N} \\
& & optic neuropathy & & & across entire vision & & \\

\bottomrule
\end{tabular}
\caption{Participant demographics and visual conditions in \colorchange{the outdoor navigation study} (Study II).}
\Description{This table lists out the demographic information and visual conditions of the 13 participants in the outdoor navigation study (Study 2).}
\label{tab:demographics_outdoor}
\end{table}


\end{document}